%
\magnification=\magstep1
\nopagenumbers \headline={\tenrm\hfil --\folio--\hfil}
\baselineskip=18pt  \lineskip=3pt minus 2pt   \lineskiplimit=1pt
\hsize=15true cm  \vsize=24true cm 
\def\tomb{\phantom{.}\hfill\vrule height.4true cm width.3true cm \par
\smallskip\noindent}
\def\ker{{\rm Ker\,}}  \def\chop{\hfill\break}  
\def\f #1,#2.{\mathsurround=0pt \hbox{${#1\over #2}$}\mathsurround=5pt}
\def\s #1.{_{\smash{\lower2pt\hbox{\mathsurround=0pt $\scriptstyle #1$}}
\mathsurround=5pt}}
\def\r{{\hbox{\mathsurround=0pt$\rm I\! R$\mathsurround=5pt}}}
\def\mapdownl #1;{\vcenter{\hbox{$\scriptstyle#1$}}\Big\downarrow}
\def\mapdownr #1;{\Big\downarrow\rlap{$\vcenter{\hbox{$\scriptstyle#1$}}$}}
\tolerance=1600 \mathsurround=5pt  
\def\maprightu #1;{\smash{\mathop{\longrightarrow}\limits^{#1}}}
\def\maprightd #1;{\smash{\mathop{\longrightarrow}\limits_{#1}}}

\def\brc #1,#2.{\left\langle #1\,|\,#2\right\rangle}
\def\rn#1{{\romannumeral#1}}  \def\cl #1.{{\cal #1}}
\def\convl #1,#2.{\mathrel{\mathop{\longrightarrow}\limits^{#1}_{#2}}}
\def\convr #1,#2.{\mathrel{\mathop{\longleftarrow}\limits^{#1}_{#2}}}
\def\set #1,#2.{\left\{\,#1\;\bigm|\;#2\,\right\}}
\def\theorem#1@#2@#3\par{\smallskip\parindent=.6true in \itemitem{\bf #1}
{\sl #2}\parindent=20pt\smallskip\itemitem{\it Proof:\/}#3\tomb}
\def\thrm#1"#2"#3\par{\smallskip\parindent=.6true in \itemitem{\bf #1}
{\sl #2}\parindent=20pt\smallskip\itemitem{\it Proof:\/}#3\tomb}
\def\teorem#1@#2@ {\smallskip\parindent=.6true in \itemitem{\bf Theorem #1}
{\sl #2}\hfill \parindent=20pt\smallskip\noindent} 
\def\fteorem#1@#2@ {\smallskip\parindent=.6true in \itemitem{\bf #1}{\sl #2}
\hfill\parindent=20pt\smallskip\noindent}    
     
\def\shift #1;{\mathord{\phantom{#1}}}  
\def\ref #1.{\mathsurround=0pt${}^{#1}\phantom{|}$\mathsurround=5pt}
\def\cross #1.{\mathrel{\jackup 3,\mathop\times\limits_{#1}.}}
\def\jackup#1,#2.{\raise#1pt\hbox{\mathsurround=0pt $#2$\mathsurround=5pt}}
\def\hlf{{\f 1,2.}}

\def\bbrc #1,#2,#3.{\langle #1 |\,#2\,|#3\rangle}
   
\def\cst#1,{{C^*(#1-\un)}}
\def\un{\hbox{\mathsurround=0pt${\rm 1}\!\!{\rm 1}$\mathsurround=5pt}}

\def\alg#1.{{C^*(#1)}}

\def\ccr #1,#2.{{\overline{\Delta(#1,\,#2)}}}

\input amssym.def
\input amssym.tex
\def\r{\Bbb R}
\def\b#1'{{\bf #1}}
\def\rest{\mathord{\restriction}}
\def\dom{{\rm Dom}\,}

\pageno=1  \noindent
\centerline{\bf Constrained Dynamics for Quantum Mechanics I.}
\centerline{\bf Restricting a
Particle to a Surface.}
\vglue .3in
\centerline{Hendrik Grundling\ref 1). and C.A. Hurst\ref 2).}
\vglue 1in
\itemitem{{\bf Abstract}}{\sl
We analyze constrained quantum systems where the dynamics do not preserve
the constraints. This is done in particular for the restriction of a quantum
particle in $\r^n$ to a curved submanifold, and we propose a method of 
constraining
and dynamics adjustment which produces the right Hamiltonian  on
the submanifold when tested on known examples.
This method will be the germ of a ``Dirac algorithm for quantum
constraints.''
We generalise it to the situation where
the constraint is a general selfadjoint operator with some additional 
structures.}
\chop
[PACS classification: 03.65.Db, 02.30.Tb, 02.40.Vh]
\par\noindent
\vglue 2in
\hrule
\centerline{\ref 1). Department of Pure Mathematics, University of
New South Wales,}
\centerline{ P.O. Box 1, Kensington, NSW 2033, Australia.}
\centerline{ email: hendrik@maths.unsw.edu.au}
\centerline{\ref 2). Dept. Physics and Mathematical Physics, University
of Adelaide,}
\centerline{GPO Box 498, Adelaide, SA 5001, Australia.}
\centerline{email: ahurst@physics.adelaide.edu.au}
\vfill\eject

\beginsection 1. Introduction.

The problem of enforcing constraints in classical mechanics has 
had satisfactory solutions
for some time now [Di, GNH, MW].
The same is not true in quantum mechanics, contrary to claims
of ``quantizations'' of classical constrained systems after
constraining.
Quantization maps are rarely unique nor well-defined [VH,
Wo, Ri, GGT]. Moreover, 
using noncommutativity, some
constraints can be defined in quantum mechanics with zero classical
limit. This means that the general problem of quantum constraints 
needs to be solved in the quantum arena  without appealing to
classical methods.

There is presently a wide variety of methods for imposing quantum
constraints [Di, GH, HT, La, 
SW]
but these methods deal with the kinematics only at the
quantum level, and their interrelations
are unclear. 
When the given quantum dynamics preserves the constrained
subsystem that is fine, one only needs to restrict it to the
constrained subsystem to obtain the constrained dynamics.
However, when this is not the case
 the problem arises of how 
to appropriately adjust the constraint set and the dynamics into
a stable system. (New constraints produced by such a method
are generally called secondary constraints).
The obvious strategy of fixing the dynamics and extending the constraint
set to its orbit under the dynamics, produces the wrong physics in
examples. It is our opinion that the question of secondary quantum 
constraints and dynamics adjustment is a physical problem,
and cannot be decided by mathematical arguments alone.
In this paper we set out to solve this problem of secondary
quantum constraints in the limited class of systems consisting
of the restriction of a quantum particle on $\r^n$ to
a smooth (possibly curved) submanifold i.e. the case of holonomic
constraints (cf. p75 [Ar]).
The motivation for this choice is as follows:
\item{(1)} there are several examples of quantum mechanics on surfaces 
available (e.g. ${\bf S}^2$) against which we can test the results of 
our analysis.
\item{(2)} in $\r^n$ there is ample geometry available to guide
the intuition, e.g. a metric, hence normal vectors and Lebesgue measure,
\item{(3)} generalisations are easy from this class,
in fact in Sect. 6 we propose a generalisation of
 it to the case of a general selfadjoint 
operator on a Hilbert space with additional structures,
\item{(4)} a constraint in this situation is of the form $\varphi(q)=0$
which has an immediate obvious quantization in the Schrodinger
representation, whereas a constraint
of the form $\varphi(q,\,p)=0$ need not have a unique quantization,
\item{(5)} physically these systems in $\r^3$ can be approximated by very thin
films or wires and so are close to experimental verification. 
\item{(6)} the exposition of constraint methods for these systems
are more transparent than for more general ones,
\item{(7)} holonomic quantum constraints have been considered in the physics
literature [Ma, Fu].

\noindent The kinematics of quantum systems of this type has
already been done concretely by Landsman [La] and abstractly
at the C*--level by Grundling and Hurst [GH], so any
new proposed method should be compatible with these. 
We will not consider BRST--methods [HT], as we believe
these not to be equivalent at the quantum level
to the other methods [Gr, LL, Mc].
 So to summarize;
we aim to solve for this class of systems the problem of secondary
 quantum constraints, i.e. given a constrained
system where the dynamics do not preserve the constraints, to find
a method of adjustment which agrees on known constrained systems
with the right physics.

The architecture of the paper is as follows; in
Sect. 2 we summarize the Dirac procedure for constraining a
classical particle to a submanifold, both local and global,
and we also do two examples.
In Sect. 3 we consider the question of quantum constraints, we first
summarize the usual Dirac procedure, then start to analyze the 
situation of restricting a quantum particle in $\r^n$ to
a lower dimensional submanifold $\Gamma$. We obtain the constraining map
$\kappa$ from the Hilbert space $L^2(\r^n)$ of the unconstrained particle
to that of the constrained particle $L^2(\Gamma)$, and we discuss
generalisations to other types of constrained systems.
In Sect. 4 we solve the problem of how to obtain from the Hamiltonian of
the original particle a constrained Hamiltonian on $L^2(\Gamma)$, and we apply
it to two examples;- restricting a free particle in $\r^3$ to a sphere and
a cylinder, and we obtain in each case the correct Hamiltonian.
In Sect. 5, we consider how to obtain observables on $L^2(\Gamma)$
from the original observables on $L^2(\r^n)$, i.e. we analyse 
when we obtain sensible liftings of operators through the map $\kappa$,
and how to obtain a suitable field algebra for the constrained particle.
In Sect. 6 we suggest a generalisation of the method of the preceding
sections to the constraining situation where the constraint is a selfadjoint
operator on an abstract Hilbert space, with zero in the continuous
part of the spectrum. We find that we need to assume some additional
structure to do the work which the geometry of $\r^n$ does in the 
method of the preceding sections.
The reader in a hurry can start with Sect. 3.

\beginsection 2. Classical Secondary Constraints.

We start by recalling the basic Dirac--Bergman method [Di, Su,
SM] which is of course a local procedure.
However to keep this brief, we only present it in the 
context of a particle in $\r^n$ constrained to an $(n-1)\hbox{--dimensional}$
submanifold $\Gamma$, which is the system at the focus of this paper.
\def\u #1'{{\underline{#1}}}

The full phase space is $\r^{2n}$ with generic point
$(\b q',\,\b p')$ and the usual Poisson algebra $\cl P.=
\big(C^\infty(\r^{2n}),\,\{\cdot,\cdot\}\big)$ is the
setting for the problem.
We specify a smooth $(n-1)\hbox{--dimensional}$ submanifold
$\Gamma$ in $\r^n$  as the zero set of a smooth
function $\varphi\in C^\infty(\r^n)$, i.e. $\Gamma=\varphi^{-1}(0)$.
We also assume $\nabla\varphi
\not=0$ on $\Gamma$ so we can use the gradient to define a normal
on $\Gamma$.
The Hamiltonian for a particle in a 
potential $V$ is:
$$H_c={1\over 2m}\,\b p\cdot p'+V(\b q')\;.$$
So we have a constrained system with a single primary constraint 
$\varphi(\b q',\,\b p'):=\varphi(\b q')$
and primary constraint manifold in phase space
$\varphi^{-1}(0)=\Gamma\times\r^n=:\Gamma_p$.
The Dirac procedure starts with the new Hamiltonian:
$$H_p:={1\over 2m}\b p\cdot p'
+V(\b q')+\mu\varphi\;,$$
where $\mu\in\cl P.$ is a free multiplier, to be determined below.
So the new time evolution for $A\in\cl P.$ on $\Gamma_p$ is
$$\dot A\;\rest\,\Gamma_p=\{A,\,H_p\}\;\rest\,\Gamma_p
=\Big(\{A,\,H_c\}+\mu\cdot\{A,\,\varphi\}\Big)\rest\Gamma_p\,.$$
Since time evolution must preserve $\Gamma_p$ we require that
$$0=\dot\varphi\rest\Gamma_p=\{\varphi,\, H_p\}\rest\Gamma_p
=\f 1,m.\b p'\cdot(\nabla\varphi)\rest\Gamma_p\,.$$
Thus $\chi:=\f 1,m.\b p'\cdot(\nabla\varphi)$ is a secondary constraint
determining a smaller submanifold
$\Gamma_f:=\varphi^{-1}(0)
\cap\chi^{-1}(0)$, which again must be preserved by time evolution.
In the examples in the rest of this paper we will constrain free motion
to a surface, so for this case $(V=0)$: 
$$\leqalignno{\dot\xi\rest\Gamma_f&= 0=\Big(\{\xi,\,H_c\}
+\mu\cdot\{\xi,\,\varphi\}\Big)\rest\Gamma_f\;,\quad\hbox{where}\quad
\xi\quad\hbox{denotes}\quad\chi,\;\varphi\cr
0&=\Big(\{\f 1,m.\b p'\cdot(\nabla\varphi),\,\f 1,2m.\b p\cdot p'\}
+\mu\{\f 1,m.\b p'\cdot(\nabla\varphi),\,\varphi\}\Big)\rest\Gamma_f
&\hbox{so}\cr
&=\Big({1\over 2m^2}\b p'\cdot\{\nabla\varphi,\,\b p\cdot p'\}
+\f 1,m.\mu(\nabla\varphi)\cdot\{\b p',\,\varphi\}\Big)\rest\Gamma_f\cr
&=\Big({1\over m^2}\b p'\cdot\big(\nabla(\b p'\cdot\nabla\varphi)\big)-
\f 1,m.\mu(\nabla\varphi)\cdot(\nabla\varphi)\Big)\rest\Gamma_f\,.\cr}$$
Since $\nabla\varphi\not=\b 0'$ on $\Gamma$, we solve on $\Gamma_f$:
$$\leqalignno{\mu&={\b p'\cdot\big(\nabla(\b p'\cdot\nabla\varphi)\big) \over
m\|\nabla\varphi\|^2}={(\b p'\cdot\nabla)^2\varphi\over
m\|\nabla\varphi\|^2}\cr
H_p&=\f 1,2m.\b p\cdot p'+\varphi\cdot {(\b p'\cdot\nabla)^2\varphi
\over m\|\nabla\varphi\|^2}\;.&\hbox{thus}\cr}$$
Notice that the second term in $H_p$ when extended near $\Gamma_f$
acts like a ``potential'' to keep the motion on $\Gamma$,
and that the secondary constraint
$$\chi=\big(\f 1,m.\b p'\cdot\nabla\varphi\big)\rest\Gamma=0\eqno{(2.2)}$$
guarantees that the motion is tangent to the surface and
there are no more constraints. Moreover, since
$$\{\varphi,\,\chi\}\rest\Gamma_f=\f 1,m.\|\nabla\varphi\|^2\rest\Gamma_f
\not=0\;,$$
the constraints are second class.
The Dirac algorithm continues to construct a Dirac bracket, but we will not need 
this. For later use we specialize the above to a sphere and a cylinder.

Let $\Gamma$ be a sphere of radius $a$ in $\r^3$. 
The phase space is  $\r^6$,
$\Gamma_p=\b S'^2\times\r^3$ and we have
$$\varphi(\b q')=\b q\cdot q'-a^2\;,\quad\hbox{so}\quad
\chi=\f 1,m.\b p'\cdot\nabla\varphi=\f 2,m.\b p\cdot q'$$
and since $\nabla(\b p'\cdot\nabla\varphi)=2\b p'$  here, we get
$$H_p={1\over 2m}\b p\cdot p'\Big(1+{\b q\cdot q'-a^2\over
\|\b q'\|^2}\Big)\eqno{(2.3)}$$
An explicit calculation produces
$$H_p\rest\Gamma_f = \f 1,2m.\b p\cdot p'\Big|_{|\b q'|=a\atop\b q\cdot p'=0}
={|\b L'|^2\over 2ma^2}
\eqno{(2.4)}$$
where $\b L'=\b q'\times\b p'\Big|_{r=a}$ is angular momentum
on the sphere.\chop

If we take $\Gamma$ to be a cylinder of radius $a$ around the $z\hbox{--axis}$
in $\r^3$, we have
$$\eqalignno{\varphi(\b q')&=q_1^2+q_2^2-a^2\;,\qquad\hbox{so}\qquad
\chi=\f 2,m.(p_1q_1+p_2q_2)\cr
H_p&=\f 1,2m.\b p\cdot p' + (q_1^2+q_2^2-a^2){p_1^2+p_2^2\over 2m
(q_1^2+q_2^2)}\,.&(2.5)\cr}$$
And now:
$$H_p\rest\Gamma_f=\f 1,2m.\b p\cdot p'\Big|_{q_1^2+q_2^2=a^2\atop 
p_1q_1=-p_2q_2}
={1\over2m}\Big({L_3^2\over a^2}+p_3^2\Big)\rest\Gamma_f\eqno{(2.6)}$$
where $L_3:=q_1p_2-q_2p_1$ is the usual angular momentum around the
$q_3\hbox{--axis}$.\chop

\medskip
Next we recall the global version of the Dirac--Bergman method of constraints
as worked out by Gotay, Nester and Hinds [GNH], 
which is a considerable simplification.
Start with a symplectic manifold $(M,\,\omega)$ (finite dimensional here)
and a Hamiltonian $H\in C^\infty(M)$. Let $M_1\subset M$ be a given
primary constraint submanifold. Now the evolution equation $\iota\s X.\omega=dH$
may not have solutions for the vector field $X$ tangential to $M_1$, so let
$$M_2:=\set m\in M_1,dH(m)={\iota\s X.}\omega(m)
\quad\hbox{for some}\; X\in T_mM_1.\,.$$
However the time evolution must now preserve $M_2$, so iterate:
$$M_{k+1}:=\set m\in M_k,dH(m)={\iota\s X.}\omega(m)\quad\hbox{for some}\;
X\in T_mM_k.\;.$$
When this iteration converges sensibly, we obtain a final constraint 
manifold $M_c$ on which $dH=\iota\s X.\omega$ has solutions 
$X\in\Gamma^\infty(M_c)\equiv\hbox{smooth}$vector fields on $M_c$,
which provide the desired time evolutions.
Geometrically we can think of $M_c$ as the largest submanifold
which has a vector field $X$ which is tangential at each 
$m\in M_c$ to a trajectory of the time evolution of the
original unconstrained manifold $M$. The completeness of $X$
is still an open question for the general analysis.
(Below we will refer to the above as the GNH--algorithm).

We will take this geometric picture as the guiding principle
for constraining the dynamics of a quantum system below
in Sect. 4.

\beginsection 3. Constraining the representation space.

For reference we recall the usual Dirac prescription
for first class quantum constraints. Given a field algebra
$\cl A.$ preserving a dense domain $\Delta$ in a Hilbert space
$\cl H.$, as well as a Hamiltonian $H\in\cl A._{sa}$ and
a set of constraints $\set \hat\varphi_i\in\cl A._{sa},i\in I.$
(where $I$ is an abstract index set),
one selects the physical subspace by
$$\cl H._{\rm phys}:=\overline{\set\psi\in\Delta,\hat\varphi_i\psi=0\;\;
\forall\, i.}$$
and this is assumed to be nonzero (hence zero must be in the discrete 
spectrum of each $\hat\varphi_i$). Decompose
$\cl H.=\cl H._{\rm phys}\oplus\cl H._{\rm phys}^\perp$,
so for each $A\in\cl A.$ we have a decomposition
$A=\left(\matrix{a&b\cr c&d\cr}\right)$.
Clearly the only $A\in\cl A.$ which, together with $A^*$ can be restricted
to $\cl H._{\rm phys}$ are those of the form
$A=\left(\matrix{a&0\cr 0&d\cr}\right)$. Denote the *--algebra of these by
$\cl O.$. Those which restrict trivially are of the form
$A=\left(\matrix{0&0\cr 0&d\cr}\right)$ and comprise an ideal
$\cl D.$ of $\cl O.$. Then the *--algebra of physical observables are
$$\cl O.\rest\cl H._{\rm phys}\cong\cl O./\cl D.=:\cl R.\,.$$
If $\exp itH$ preserves $\cl H._{\rm phys}$ then
${\rm Ad}\,\exp(itH)$ lifts to automorphisms on $\cl R.$.
Otherwise we need to alter it (cf. Sect. 4).
\def\P_0{P_{\rm phys}}
Let $\P_0$ be the projection on $\cl H._{\rm phys}$ and define the 
sesquilinear form
$$(\psi_1,\,\psi_2)_D:=(\P_0\psi_1,\,\P_0\psi_2)=(\psi_1,\,\P_0\psi_2)
\eqno{(3.1)}$$
then $\ker(\cdot,\cdot)_D=\cl H._{\rm phys}^\perp$ and we write
$\P_0$ as a factorization $\P_0:\cl H.\to\cl H./\ker(\cdot,\cdot)_D
\cong\cl H._{\rm phys}$ (obvious identification),
and we can think of $\cl O.$ as all $A\in\cl A.$ which
together with $A^*$ can lift through the factorisation.

Next we introduce our main object of study.
Consider a quantum particle in $\r^n$ in the Schr\"odinger representation.
Its  basic data consists of the
Hilbert space $\cl H.=L^2(\r^n,\,\mu)$ (with $\mu$ the Lebesgue
measure), and on the dense subspace
$C_c^\infty(\r^n)$ of smooth functions of compact support, we have the 
position operator $(\hat\b q'\psi)(\b x'):=\b x'\psi(\b x')$,
the momentum operator $(\hat\b p')(\b x'):=(i\nabla\psi)(\b x')$
and a Hamiltonian $H$ (which is $\f 1,2m.\hat\b p'\cdot\hat\b p'$
 when the particle
is free). These operators are all essentially selfadjoint and
preserve $C_c^\infty(\r^n)$.
We wish to constrain this particle to a given
 smooth $(n-1)\hbox{--dimensional}$
submanifold $\Gamma\subset\r^n$.
Assume we have a bounded real--valued constraint function
$\varphi\in C^\infty(\r^n)$ with $\nabla\varphi$ nonzero
and bounded on a neighbourhood 
of $\Gamma$, and
$$\Gamma=\set \b x'\in\r^n,\varphi(\b x')=0.=\varphi^{-1}(0)\;.$$
(Notice that the level hypersurfaces of $\varphi$ form a foliation
of this neighbourhood [To].)
Quantize $\varphi$ by the multiplication operator $\varphi(\hat\b q')$, i.e.
$$(\hat\varphi\psi)(\b x')=\varphi(\b x')\cdot\psi(\b x')\;\;\forall\,\psi
\in \cl H.$$
which is selfadjoint since $\varphi$ is real--valued, and bounded because
$\varphi$ is.\chop
Assume we are given a concrete field algebra $\cl F.\subset\cl B(H).$,
which should be a C*--algebra containing all relevant operators in
bounded form, e.g.
$\hat\varphi$, $\exp i\b a'\cdot\hat\b q'$, $\exp i\b a'\cdot\hat\b p'$
($\b a'\in\r^n$). The choice of field algebra will turn out to be
important for producing a meaningful algebra of observables in
the constrained system, but we will return to this matter in a later
section.
In summary, we are assuming the following data:
\item{$\bullet$} the operators $\hat\b p'$, $\hat\b q'$ and $H$ on 
$C_c^\infty(\r^n)\subset L^2(\r^n)$,
\item{$\bullet$} A bounded $\varphi\in C^\infty(\r^n)$ 
with $\Gamma=\varphi^{-1}(0)$
a smooth $(n-1)\hbox{--dimensional}$ submanifold, and with $\nabla\varphi$
nonzero and bounded in a neighbourhood of $\Gamma$,
\item{$\bullet$} A unital field algebra $\cl F.$ on $L^2(\r^n)$.

\noindent
We would like to find a way of imposing the constraint ``$\hat\varphi=0$''
on this system.
However since $\Gamma$ is a null--set of $\mu$, we have $\ker\hat\varphi
=\{0\}$, and so Dirac's method of restriction to the kernel of the
constraint fails. The fact that zero is in the continuous spectrum of
$\hat\varphi$ is physically due to the uncertainty principle;-- 
a particle confined to $\Gamma$ will violate Heisenberg's uncertainty
principle in a direction normal to $\Gamma$.

At present there are two methods which can handle the above situation;--
the T--procedure of [GH], as well as Landsman's
application of Rieffel induction [La]. 
The T--procedure ``ignores'' the original representation of
$\cl F.$, considers $\cl F.$ as an abstract C*--algebra, for which it then
finds those representations $\pi$ for which zero is in the 
discrete part of the spectrum of $\pi(\hat\varphi)$ and works out the 
algebraic structures associated with factoring $\ker\pi(\hat\varphi)$
out of each of those representations.
On the other hand, Landsman's method assumes a locally compact group
$H$ with a continuous proper action
on $\r^n$ such that $\Gamma$ is precisely the
set of points left invariant.
Assuming some initial representation $\pi$ of $H$ on 
$L^2(\r^n)$, he makes
$C_c(\r^n)$ into an $(\cl A.\mathord{-}\cl B.)\hbox{--bimodule}$
where $\cl B.$ is the algebra $C_c(H)$ (with convolution for multiplication)
and $\cl A.$ is the part of the commutant of $\pi(H)$ preserving
$C_c(\r^n)$. There is then a natural rigging map, and this allows
a Rieffel induction from the trivial representation of $\cl B.$
to $\cl A.$. 

The T--procedure selects the set of all representations in which
the Dirac algorithm makes sense, which necessarily excludes the
original representation of the current system.
This means that one loses the original
physical interpretation of the state vectors, and one
has the problem of which representation to choose for the 
constrained system. From the point of
view of the physics, one may sometimes be more interested in
constructing the constrained operators directly out of the
original ones.
Landsman's method does this, but requires some additional group
structure, and quantizes a fairly small algebra.
Both methods do only the kinematics, and if the dynamics do not
preserve $\Gamma$ it is hard to see from these structures what
should be done.

We will propose a method for constraining
the kinematics of a quantum particle to $\Gamma$
which generalises the usual Dirac prescription
and in which
the constrained operators are explicitly constructed from the
original ones. It will then be fairly easy to see what to do
with the dynamics.

In a sense the solution of the kinematics problem is already known.
One takes the subspace 
$C_c(\r^n)\subset L^2(\r^n)$ and restricts it to $\Gamma$,
thus obtaining a dense subset of the physical Hilbert space
$L^2(\Gamma,\,\gamma)$ ($\gamma$ denotes the measure induced on
$\Gamma$ by $\mu$). The restriction is the same as factoring out 
the subspace
$$N:=\set f\in C_c(\r^n),f\,\rest\Gamma=0.$$
from $C_c(\r^n)$. The observables are those operators which ``restrict
to $\Gamma$,'' i.e.. operators $A$ which preserve $C_c(\r^n)$ and $N$,
in which case $A$ lifts through the restriction to define an operator
on $C_c(\Gamma)\subset L^2(\Gamma)$. This is the procedure which we
intend to refine.
 Note however that the restriction map
$\rho:C_c(\r^n)\to C_c(\Gamma)$ is unbounded with respect to the 
$L^2\hbox{--norms}$ and it is not closable as an operator
(let $\psi_k\in L^2(\r^n)$ be a sequence of continuous functions
with compact support,
all of which have the same restriction to a nonzero $\theta
\in L^2(\Gamma)$, and which converge to zero w.r.t. the $L^2
\hbox{--norm}$.
Then $\psi_k\to 0$, $\rho(\psi_k)\to\theta\not=0$, so
$\lim\rho(\psi_k)\not=\rho(\lim\psi_k)$).
So we expect some pathologies to arise in operator questions.
We would also like to build the inner product of
$L^2(\Gamma)$ explicitly out of the inner product of $L^2(\r^n)$.
To do this, we adapt a well--known argument from
statistical mechanics [Kh], and p80 of [Fa].

Since $\nabla\varphi(\b x')\not=0$ on a neighbourhood of $\Gamma$,
the critical points of $\varphi$ are a finite distance away from
$\Gamma$. So we have the geometrically obvious:
\thrm Lemma 3.2."
There is a $t>0$ such that in $\varphi^{-1}[-t,\,t]$
we have locally
 a smooth curvilinear orthogonal coordinate system $(\b y',\,\varphi)$
where $\b y'=\hbox{constant}$ are curves normal to the level
hypersurfaces of $\varphi$."
Consider the vector field
$X(\b x')=\b n'(\b x')\big/|\nabla\varphi(\b x')|$ on $\varphi^{-1}(-t,\,t)$,
where $\b n'=\nabla\varphi\big/|\nabla\varphi|$.
Then its integral curves are normal to all level surfaces of $\varphi$
in $\varphi^{-1}(-t,\,t)$, and moreover for each $s\in(-t,\,t)$ we
have a diffeomorphism $\alpha_s:\varphi^{-1}(0)\to\varphi^{-1}(-t,\,t)$
such that $\alpha_0=\iota$ and $\f d,ds.\alpha_s(\b x')=X(\alpha_s(\b x'))$
[AM].
In fact $\varphi(\alpha_s(\b x'))=s$ for all $\b x'\in\varphi^{-1}(0)
=\Gamma$, which we see as follows. For small $\varepsilon$:
$$\leqalignno{\alpha_\varepsilon(\b x')&=\b x'+\varepsilon\,
{\b n'(\b x')\over|\nabla\varphi(\b x')|}+O(\varepsilon)\cr
\varphi(\alpha_\varepsilon(\b x'))&=\varphi(\b x')+
\big(\varepsilon\big/|\nabla\varphi(\b x')|\big)\,\nabla\varphi(\b x')\cdot
\b n'(\b x')+O(\varepsilon)&\hbox{so}\cr
&=\varphi(\b x')+ \varepsilon +O(\varepsilon)
\qquad\forall\,\b x'\in\varphi^{-1}(-t,\,t)\cr}$$
using a Taylor expansion and $\b n'=\nabla\varphi\big/
|\nabla\varphi|$. So ${d\over ds}\varphi(\alpha_s(\b x'))=1$
with solution $\varphi(\alpha_s(\b x'))=s+\varphi(\b x')$
and if $\b x'\in\Gamma$, the last term is zero.
 Thus we have diffeomorphisms $\alpha_s:\varphi^{-1}(0)
\to\varphi^{-1}(s)$. Now equip an open neighbourhood $U\subset\varphi^{-1}(0)$
with local coordinates $\b y'$ by a chart, then $\alpha_s$ will
equip $\alpha_s(U)\subset\varphi^{-1}(s)$ with the same coordinates.
Thus we have the desired local coordinates $(\b y',\,\varphi)$ on 
the set $V:=\varphi^{-1}(-t,\,t)\cap\alpha\s(-t,\,t).(U)$.
(The incompleteness of $X$ on $\varphi^{-1}[-t,\, t]$ is not an
obstacle to the argument;-- just attenuate $X$ with appropriate
smooth bump functions to make it integrable with speed 1 on
$\varphi$ where we 
want it).

For a function $f\in L^1(\r^n)$ we have
$$\int_Vf(\b x')\, d\mu(\b x')=\int_Vf(\b y',\,\varphi)\,J(\b y',\,\varphi)
\,d\b y'\,d\varphi$$
where $J$ is the Jacobian and $V$ is as in the preceding
proof. Observe that for a fixed $\varphi=s$
that $J(\b y',\,s)\,d\b y'$ is not yet the surface measure $d\gamma_s$
on $\varphi^{-1}(s)$. For that the orthogonal coordinate $\varphi$
needs to be expressed in terms of the length of the curves $\b y'=
\hbox{constant}$. Since $d\varphi=|\nabla\varphi|\,d\ell$ with
$d\ell$ the length measure on a curve $\b y'=\hbox{constant}$, 
we conclude that $d\gamma_s(\b y')=[J\cdot|\nabla\varphi|](\b y',\,s)\,
d\b y'$. Now
$$\eqalignno{\int_V\big|\nabla\varphi(\b x')\big|\,f(\b x')\,d\mu(\b x')
&=\int_Vf(\b y',\,\varphi)\,[J\cdot|\nabla\varphi|]\,d\b y'\,d\varphi\cr
&=\int_{-t}^t\Big(\int_Uf\cdot J\cdot|\nabla\varphi|\,d\b y'\Big)\,
d\varphi&(3.3)\cr}$$
and this is the expression we wish to exploit. Let 
$f\in C_c(\r^n)$, and let the thickness of the
shell $\varphi^{-1}[-t,\, t]$ around $\Gamma$ approach zero, then
$$\eqalignno{\lim_{s\to 0}{1\over 2s}\int_{V\cap\varphi^{-1}[-s,\,s]}
|\nabla\varphi|\cdot f\,d\mu&=\lim_{s\to 0}{1\over 2s}\int_{-s}^s
\Big(\int_Uf\cdot J\cdot|\nabla\varphi|\,d\b y'\Big)\,d\varphi\cr
&=\int_U\big(f\cdot J\cdot|\nabla\varphi|\big)(\b y',\, 0)\,d\b y'\cr}$$
where the use of the fundamental theorem of calculus is justified because 
the function $$\varphi\to\int_U\big(f\cdot J\cdot|\nabla\varphi|\big)(
\b y',\,\varphi)\,d\b y'$$
is continous due to the uniform continuity of the integrand;-- a
consequence of $f\in C_c(\r^n)$ and $J\cdot|\nabla\varphi|\in
C^\infty(V)$. Thus
$$\lim_{s\to 0}{1\over 2s}
\int_{V\cap\varphi^{-1}[-s,\,s]}|\nabla\varphi|\cdot f\,d\mu
=\int_U(f\,\rest\,\Gamma)\,d\gamma\;.$$
By doing this for all open patches $U\subset\Gamma$ (equipped with charts),
we conclude
$$\lim_{s\to 0}{1\over 2s}\int_{\varphi^{-1}[-s,\,s]}|\nabla\varphi|\cdot f
\, d\mu=\int_\Gamma(f\rest\Gamma)\,d\gamma\eqno{(3.4)}$$
for all $f\in C_c(\r ^n)$.

In particular, let $\psi_1,\; \psi_2\in C_c(\r^n)$, then we have
$$\eqalignno{\big(\psi_1\rest\Gamma,\,\psi_2\rest\Gamma\big)\s L^2(\Gamma).
&=\lim_{t\to 0}
{1\over 2t}\int_{\varphi^{-1}[-t,\,t]}\overline{\psi_1}\cdot \psi_2\cdot|\nabla
\varphi|\, d\mu\cr
&=\lim_{t\to 0}{1\over 2t}\left(\psi_1,\,|\nabla\varphi|\cdot\chi\s\varphi^{-1}
[-t,\,t].\psi_2\right)\cr
&=\lim_{t\to 0}{1\over 2t}\left(\psi_1,\,\hat hP_t
\psi_2\right)&{(3.5)}\cr}$$
where we used the notation $(\hat h\psi)(\b x'):=|\nabla\varphi(\b x')|\psi(\b
x')$ and $P_t:=\hat\chi\s\varphi^{-1}[-t,\, t].$
and this is the desired relation between the inner products
of the initial space $L^2(\r^n)$ and of the physical space
$L^2(\Gamma)$.
(Note that $P_t$ is just the spectral projection of $\hat\varphi$
on the interval $[-t,\, t]$, and that $\hat h$ is a positive
bounded operator on $P_t\cl H.$).
Now the right hand side of this equation will exist for a much larger class
of functions in $L^2(\r^n)$ than $C_c(\r^n)$, though for some of these
the restrictions to $\Gamma$ on the left hand side may not be defined.
However the limit in (3.5) will definitely fail to exist for some
elements of $L^2(\r^n)$.
Define
$$(\psi_1,\,\psi_2)_\Gamma:= \lim_{t\to 0}{1\over 2t}
\left(\psi_1,\,\hat hP_t\psi_2\right)
\eqno{(3.6)}$$
for all pairs $\psi_1,\,\psi_2$ for which the rhs is defined and
finite. To obtain a positive sesquilinear form from (3.6),
we need to decide on a dense domain on which $(\cdot,\cdot)_\Gamma$
is defined and finite. Clearly $C_c(\r^n)$ is one such domain,
but it is not maximal. In fact, $(\cdot,\cdot)_\Gamma\rest
C_c(\r^n)$ has no closed extensions as a sesquilinear form
(cf. [RS] p278), so there is no canonical way of
getting a maximal domain. The set
$$S:=\set\psi\in L^2(\r^n),(\psi,\,\psi)_\Gamma\;\;\hbox{exists
and is finite}.$$
is not a linear space since $(\psi_1,\,\psi_2)_\Gamma$ need not
exist for $\psi_1,\,\psi_2\in S$ as one can verify with easy examples.
Nevertheless, some domains containing $C_c(\r^n)$ 
seem quite natural, e.g. the domain
$C_\varphi$ consisting of those $\psi\in L^2(\r^n)$ for which there
is some $t>0$ (depending on $\psi$) such that $P_t\psi\in
C_c\big(\varphi^{-1}[-t,\, t]\big)$, and clearly the argument which
led to eq. (3.5) is still valid for $\psi_i\in C_\varphi$.
Now $C_\varphi$ is still not maximal and
we can of course choose
a maximal defining subset for $(\cdot,\cdot)_\Gamma$ in $S$
containing $C_\varphi$, but this choice is not unique.
Equip the factor space $C_\varphi\big/\ker{(\cdot,\cdot)_\Gamma}$
with the inner product $(\cdot,\cdot)_\Gamma$, making it into a
pre--Hilbert space which we complete to obtain the Hilbert space
$\cl H._\Gamma$. We identify $\cl H._\Gamma$ with
$L^2(\Gamma)$ in a natural way:
\thrm Theorem 3.7."
With the structures above, denote by $\cl C.:C_\varphi\to
C_\varphi\big/\ker(\cdot,\cdot)_\Gamma$ the factor map, then\chop
(1)  $(\cl C.\psi_1,\,\cl C.\psi_2)_\Gamma=\big(\psi_1\rest\Gamma,
\,\psi_2\rest\Gamma\big)\s L^2(\Gamma).$ for all $\psi_i\in C_c(\r^n)$,\chop
(2)  there is a unitary $U:\cl H._\Gamma\to L^2(\Gamma,\,\gamma)$
such that  $U\cl C.\psi=\psi\rest\Gamma$ for all $\psi\in C_c(\r^n)$,
and $(U\cl C.)\big(C_c(\r^n)\big)=C_c(\Gamma)$."
(1) Since $(\psi_1,\,\psi_2)_\Gamma=(\cl C.\psi_1,\,\cl C.\psi_2)_\Gamma$,
this is already proven above.\chop
(2) Let $\psi\in C_c(\r^n)$ and define $U\cl C.\psi:=\psi\rest\Gamma$
which is in $C_c(\Gamma)$ since $\Gamma$ is a closed subset of $\r^n$.
To see that $U$ is well defined on $C_c(\r^n)\big/\ker(\cdot,\cdot)_\Gamma$,
let $\psi_1,\,\psi_2\in C_c(\r^n)$ with $(\psi_1-\psi_2)\rest\Gamma=0$.
Then by (3.4) we find $\psi_1-\psi_2\in\ker(\cdot,\cdot)_\Gamma$,
i.e. $\cl C.\psi_1=\cl C.\psi$ so $U$ is well-defined. Moreover
from (1) we see that $U$ is unitary on the dense subspace
$\cl C.[C_c(\r^n)]$ hence it extends to a unitary on $\cl H._\Gamma$.
Clearly $C_c(\r^n)\rest\Gamma=C_c(\Gamma)$, so since the image
of $U$ contains a dense subspace and U is unitary, we find
$U:\cl H._\Gamma\to L^2(\Gamma)$ onto.

This means the restriction map can be written as
$\kappa:=U\circ\cl C.:C_\varphi\to L^2(\Gamma)$, but now it implicitly
involves a limiting process, which we will exploit below. Note that 
$C_c(\r^n)\cap\ker(\cdot,\cdot)_\Gamma=\set\psi\in C_c(\r^n),
\;\psi\rest\Gamma=0.$ and that $\kappa$ 
is unbounded (w.r.t. the Hilbert space norms).

\itemitem{\bf Example.} We want to see what the above structure looks like
in the restriction of a quantum particle in $\r^3$ to a sphere. So take
$\cl H..=L^2(\r^3)$, $\hat\b p'\psi=i\nabla\psi$, $\hat\b q'\psi(\b x')
=\b x'\psi(\b x')$ as usual and for the constraint
$\varphi(\b x')=|\b x'|^2-a^2$, $a>0$. Now $(\nabla\varphi)(\b x')=2\b x'
\neq 0$ except if $\b x'=0$, and so the critical point of $\varphi$ is
distance $a$ away from $\Gamma=\varphi^{-1}(0)$. We have 
$\varphi^{-1}[-s,\, t]=\set\b x'\in\r^3,|\b x'|^2\in[a^2-s,\,a^2+t].$,
$s,\, t>0$. So
$$\eqalignno{&(\psi_1,\,\psi_2)_\Gamma=\lim_{t\to 0}{1\over 2t}
(\psi_1,\,\hat hP_t\psi_2)
\cr
&=\hlf{d\over dt}\int_{\varphi^{-1}[-t,\,t]}2|\b x'|\big(\bar\psi_1\psi_2
\big)(\b x')\,d^3x\Big|_{t=0}
\cr
&={d\over dt}\int_{\sqrt{a^2-t}}^{\sqrt{a^2+t}}r^2dr\int d
\Omega\, r\big(\bar\psi_1\psi_2
\big)(\b x')\Big|_{t=0}\cr}$$
where we used polar coordinates and $\Omega$ denotes the measure on
the unit sphere $\b S'^2$.
For functions of the form $\psi_i(\b x')=f_i(r)\,\xi_i(\theta,\,\phi)$,
$\xi_i\in L^2(\b S'^2)$, and $f_i$ continuous 
 on $(a-\varepsilon,\,a+\varepsilon)$
we have by the fundamental theorem of calculus
$$\eqalignno{(\psi_1,\,\psi_2)_\Gamma&=\Big({d\over dt}\int_{\sqrt{a^2-t}}
^{\sqrt{a^2+t}}
r^3\bar f_1f_2(r)\,dr\Big|_{t=0}
\Big)\int\bar\xi_1\xi_2(\theta,\,\phi)d\Omega\cr
&=a^2\bar f_1(a)f_2(a)
\int\bar\xi_1\xi_2\,d\Omega\,,\cr}$$
as expected from (3.5).
Moreover we have that
$\psi_i\in\ker(\cdot,\cdot)_\Gamma$ whenever 
$f_i(a)=0$.

\noindent One may ask how the method above should be
generalised to bigger classes of constraints, so towards that a few
remarks.
\itemitem{\bf Remarks.}\b (1)' If $\Gamma$ has corners (so $\varphi$ is not
smooth) we can develop an ``approximation'' to it by a sequence
of smooth submanifolds $\{\Gamma_k\}$, but need to choose a notion
of convergence for $\Gamma_k\to\Gamma$.\chop
\b (2)' If $\Gamma$ has edges we can combine the procedure above with 
a Dirac constraining. For example if we want to constrain a particle 
in $\r^3$ to the upper hemisphere $\Gamma_+$ of the sphere $\Gamma$
of the last example,
$$\Gamma_+:=\set\b x'\in\r^3,|\b x'|=a,\; x_1\geq 0.$$
then first apply the method above to the constraint
$\varphi(\b x')=|\b x'|^2-a^2$ to obtain a dense
subspace of $L^2(\Gamma)$, and follow this by applying a
Dirac procedure via (3.1) to the constraint $\zeta(\theta,\,\phi)=
\chi\s[0,\,\pi/2].(\theta)$ in $L^2(\Gamma)$.\chop
\b (3)' To constrain a particle in $\r^n$ to a smooth $(n-k)
\hbox{--dimensional}$ submanifold $\Gamma$, we need 
$k$ independent constraints $\varphi_1,\ldots,\varphi_k$
such that $\Gamma=\varphi_1^{-1}(0)\cap\ldots\cap\varphi_k^{-1}(0)$
with critical points well away from $\Gamma$. Now it is clear 
how to adapt the method based on (3.2);-- we choose a curvilinear
coordinate system $(\b y',\,\varphi_1,\ldots,\,\varphi_k)$
where $\varphi_i$ are the coordinates measured along the trajectories
of the gradient fields $\nabla\varphi_i$, and $\b y'$ are the coordinates
along the manifolds $\varphi_1^{-1}(c_1)\cap\cdots\cap\varphi_k^{-1}(c_k)$,
$\b c'$ in some $\varepsilon\hbox{--neighbourhood}$ of zero in $\r^k$.
Then with $J$ the Jacobian, the measure on $\Gamma$ is 
$d\gamma= \big(J\cdot|\nabla\varphi_1|\cdots|\nabla\varphi_k|\big)
(\b y',\,0,\ldots,\, 0)\,d\b y'$ so
(3.3) becomes for $f\in C_c(\r^n)$:
$$\lim_{\b s'\to\b 0'}[2^ks_1\cdots s_k]^{-1}
\int_{\b\varphi'^{-1}[-\b s',\b s']}
|\nabla\varphi_1|\cdots
|\nabla\varphi_k|\cdot f\,d\mu=\int_\Gamma(f\rest\Gamma)\,d\gamma$$
where $\b\varphi'^{-1}[-\b s',\,\b s']:=
\varphi_1^{-1}[-s_1,\, s_1]\cap\cdots\cap\varphi_k^{-1}[-s_k,\, s_k]$.
Further adaptations are straightforward.\chop
Alternatively, we can impose the constraints $\varphi_i$ one-by-one,
but for this we need to generalise the procedure above for $\r^n$
to general Riemannian manifolds.\chop
\b (4)' For $\varphi$ which are positive, $\Gamma=\varphi^{-1}(0)$
consists only of critical points, though if other critical points
are well away from $\Gamma$, there is still a neighbourhood of $\Gamma$
foliated by the level hypersurfaces of $\varphi$, and it is possible
to adapt the method above to this situation.\chop
\b (5)' In the case where $\varphi^{-1}(0)=\Gamma\cup\Delta$
with $\mu(\Gamma)=0\neq\mu(\Delta)$, $\Gamma\cap\Delta=\emptyset$,
neither the Dirac method (which will only produce $L^2(\Delta)$)
nor the method above seem appropriate. So what we want is a method
which will give the method above on $\Gamma$, and the Dirac
procedure on $\Delta$. (We choose not to allow surface terms
on $\Delta$). Assume the critical points of $\varphi$
are well away from $\Gamma$, and that $\Delta$ is the closure of
an open set. That is, we want a densely defined map
$\kappa:L^2(\r^n)\to L^2(\Gamma)\oplus L^2(\Delta)$. Recalling
that $L^2(\Delta)=P\s\{0\}.L^2(\r^n)$, this can be done
by defining:
$$(\psi_1,\,\psi_2)\s R.:=(\psi_1,\, P\s\{0\}.\psi_2)
+\lim_{t\to 0}{1\over 2t}(\psi^N_1,\,\hat hP_t\psi_2^N)$$
for all $\psi_i\in\b D'_R:=\set\psi\in\cl H.,
\psi^N\in C_\varphi.$ 
where $\psi^N:=\chi\s N.\cdot\psi$
and $N$ is an open neighbourhood of $\Gamma$ such that
$\overline N\cap\Delta=\emptyset$. Then proceed as before,
taking $\b D'_R/\ker(\cdot,\cdot)_R$ with $\kappa$ the
composition of the factor map and the unitary from 
$\b D'_R/\ker(\cdot,\cdot)_R$ to $L^2(\Gamma)\oplus
L^2(\Delta)$.\chop
\b (6)' At the cost of complicating the current exposition,
we can enlarge the domain $C_\varphi$ of $\kappa$
considerably. For instance, by splitting the limit (3.6) into
the average of the limits from above and below in $t$, 
and letting the domain consist of those $\psi\in L^2(\r^n)$
for which $\varphi\to\psi(\b y',\,\varphi)$ is continuous
from above and below at zero (where we used the curvilinear
coordinates of Lemma 3.2), we obtain another useful
domain for $\kappa$. We chose not to do this, because 
the additional analytic details would have obscured the
simplicity of our central idea.\chop
\b (5)' The methods proposed here for dynamics
reduction can be extended to other holonomic constraints, i.e. if
we are given a constraint $\varphi(\b q',\,\b p')$ which involves only
one member of each canonical pair $(q_i,\, p_i)$, then
a partial Fourier transform can convert it to the type
$\varphi(\b q')$ considered here.
So there is a unitary transformation of such a system to one of
the present type.

\def\S{\Bbb S}
\def\un{\Bbb I}

\beginsection 4. Constraining the Dynamics.

In this section we continue the analysis of the problem of constraining a
quantum particle in $\r^n$ to a subset
and address the problem of how to constrain the dynamics,
i.e. how to construct out of the given time evolution
on $L^2(\r^n)$ an acceptable time evolution on the physical 
Hilbert space.
We consider four cases:
\item{(1)} For an ordinary Dirac constraining we
restrict for example 
a quantum particle in $\r^n$ with Hamiltonian $H$ to live
on a set $T$ which is the closure of an open set $S$ (hence $\mu(T)\not=0$).
Kinematically, one constrains via the projection $P_{\rm phys}:
L^2(\r^n)\to L^2(T)$, as described at the start of Sect. 3. 
If the dynamics preserve $T$, i.e. each $U_t:=\exp itH$ is of the form
$U_t=\left(\matrix{a&0\cr 0& b\cr}\right)$ with respect to the
decomposition $L^2(\r^n)=L^2(T)\oplus L^2(T)^\perp$, then
there is no problem;- one just restricts $U_t$ to $L^2(T)$
to obtain the  constrained dynamics.
In practice this is a rare occurrence.
\item{(2)} For the case of a Dirac constraining as in (1) where
the dynamics does {\it not} preserve $T$,
e.g. if the particle is free, $H=\hat\b p'^2/2m$
and $T$ is compact, then $H\rest L^2(T)$ is not selfadjoint.
To equip the constrained particle with a time evolution, we need to
choose some selfadjoint extension of the symmetric operator
$H\rest C_c^\infty(S)$ (for this to make sense we need
to assume that $H$ preserves $C_c^\infty(S)$, which will be true if
it is a differential operator).
This amounts to the choice of boundary conditions, 
i.e. deciding how the particle should behave at the walls
(e.g. reflection), and this is a physical choice which cannot be
determined from mathematical considerations alone.
\item{(3)} For
the problem of the last section where we constrain
a particle in $\r^n$ with Hamiltonian $H$
to a lower dimensional submanifold $\Gamma$,
we will need to assume that $H$ is ``smooth enough'' near
$\Gamma$, i.e. $H\big({\rm Dom}\,H\cap C_\varphi\big)\subset C_\varphi$
and $\kappa\big({\rm Dom}\,H\cap C_\varphi\big)$ is dense in $L^2(\Gamma)$.
In the case where $H$ ``preserves'' $\Gamma$, i.e. $H$
preserves ${\rm Dom}\,H\cap\ker\kappa$, $H$ will clearly lift
through $\kappa$ and we only need to define the constrained 
Hamiltonian $H_\Gamma$ on $L^2(\Gamma)$ by
$H_\Gamma\kappa(\psi):=\kappa(H\psi)$ for all $\psi\in
{\rm Dom}\,H\cap C_\varphi$, and require that it is essentially
selfadjoint.
\item{(4)} The case of constraining a particle as in (3) but
where $H$ does not preserve $\Gamma$, will be our object of
study for the rest of this section. By analogy with case (2),
we expect that some physical choices will need to be made.
When $H$ is a Schr\"odinger operator with smooth potential,
physicists already know what to do;- one lets the constrained
Hamiltonian be ${1\over 2m}\Delta_\Gamma+ V\rest\Gamma$
where $\Delta_\Gamma$ is the Laplacian on $\Gamma$.

\noindent
Henceforth we assume  that $H$ on $L^2(\r^n)$ does not preserve $\Gamma$
in the sense of (3) above.
Recalling that $\kappa:C_\varphi\to L^2(\Gamma)$ involves a factorisation,
it is natural to look for linear sections of $\kappa$ which one
can use to construct liftings of operators on $C_\varphi$ to $L^2(\Gamma)$.
The image of such a section deserves a name of its own:
\item{\bf Def.} A {\it transverse space}
is a linear space
 $\cl T.\subset C_\varphi$
 such that $\ker\kappa\cap\cl T.=\{0\}$ and
$\kappa(\cl T.)$ is dense in $L^2(\Gamma)$.

\noindent Transverse spaces are abundant, and 
below we will use physical
arguments to make a choice of a transverse space but first we
show how one can use a transverse space to constrain $H$:-  given
a transverse space $\cl T.\subset\dom H\cap C_\varphi$ 
such that $H\cl T.\subset C_\varphi$, we
have the options:
\item{$(\rn1)$} Define the constrained Hamiltonian $H_{(\rn1)}$
with dense domain $\kappa(\cl T.)$ by
$$H_{(\rn1)}\kappa(\psi):=\kappa(H\psi)\,,\quad
\forall\,\psi\in\cl T..$$
This is well-defined since by definition of transverse spaces, 
$\kappa$ is injective on $\cl T.$.
\item{$(\rn2)$} Given that the choice of $\cl T.$ has some physical content
(see below),
one may object to $(\rn1)$ in that $H$ is not forced to preserve $\cl T.$
in any sense. If we want to include such a restriction, since it is
only the behaviour near $\Gamma$ which should be important, we can restrict
$H$ to $\cl T.$ ``in the limit,'' i.e. we consider the projection
$P^t_\cl T.H$ as $t$ goes to zero in the definition
of $\kappa$, where $P^t_\cl T.$ denotes
 the projection onto $P_t\overline\cl T.$.
So we propose an alternative constrained Hamiltonian $H_{(\rn2)}$ on
$\kappa(\cl T.)$ by
$$(H_{(\rn2)}\kappa(\psi),\,\kappa(\phi))\s L^2(\Gamma).
=\lim_{t\to 0}{1\over 2t}\big(P^t_\cl T.H\psi,\,\hat hP_t\phi)$$
for all $\psi,\;\phi\in\cl T.$.

\noindent Below we will test both methods for a suitable
choice of transverse space. This choice is the issue we now
want to address.
Recall that in the classical global
method of Gotay, Nester and Hinds [GNH], the idea was to adjust
the dynamics so that the motion is always tangential to $\Gamma$,
thus forcing the particle to remain on $\Gamma$.
We look for some quantum mechanical version of this.

First we want to give meaning to the concept of
motion ``tangential to $\Gamma$'' for a quantum particle.
Whilst the classical state of a particle is a point in phase space,
quantum mechanically
a state is here a vector $\psi$ in $L^2(\r^n)$.
Classically, a particle with constant mass at position $\b q'$ moves
tangentially to $\Gamma$ if its momentum $\b p'=m\dot\b q'$ is tangential
to $\Gamma$, i.e. $\b p'\in T_\b q'\Gamma$. Inspired by this, in a
quantum mechanical setting,
we say a particle in a state $\psi\in C_c^\infty(\r^n)$
moves tangentially to $\Gamma$ if $(\hat\b p'\psi)(\b q')\in
T_\b q'\Gamma$ for all $\b q'\in\Gamma$. Now, given the normal vectors
$\nabla\varphi$ to $\Gamma$, we have $\b v'\in T_\b q'\Gamma$
iff $\nabla\varphi(\b q')\cdot\b v'=0$. So a particle in a state
$\psi\in C_c^\infty(\r^n)$ moves tangentially to $\Gamma$
when $\left(\nabla\varphi\cdot\hat\b p'\psi\right)(\b x')=0$ for all
$\b x'\in\Gamma$, i.e. iff the component of the momentum of $\psi$
normal to $\Gamma$ is zero. We would like to generalise this notion
away from $C_c^\infty(\r^n)$. Since $\Gamma$ is of $\mu\hbox{--measure}$
zero, we cannot in general specify a property for an $L^2\hbox{--function}$
$\psi\in L^2(\r^n)$ on $\Gamma$. Recall that $\kappa$ is a limiting 
procedure on the shells $\varphi^{-1}[-t,\, t]$, in which eventually
$t\to 0$. So it may be enough to define tangentiality on the level sets
of $\varphi$ in such a shell, and depend on $\kappa$ to restrict 
the property to $\Gamma$. Fix a shell $S_{t_0}:=
\varphi^{-1}[-t_0,\,t_0]$ around
$\Gamma$, then we say a $\psi\in {\rm Dom}\left(\hat\b p'\rest L^2(S_{t_0})
\right)$ has momentum tangential to the level sets of $\varphi$
iff $\nabla\varphi\cdot\hat\b p'\psi=0$. Since in terms of the
local coordinate system $(\b y',\,\varphi)$ of Sect. 3 this means
$\displaystyle{{\partial\psi\over\partial\varphi}}=0$ on $S_{t_0}$, we see 
that $\psi\rest S_{t_0}$ must be
 constant in the normal direction (i.e. along the 
trajectories of the vector field $\nabla\varphi$). So $\psi\rest S_{t_0}$ is
uniquely determined by its restriction to $\Gamma$.
This
defines the notion of ``states with momentum
tangential to the level sets of $\varphi$ in $S_{t_0}$.''

Conversely, given {\it any} $\psi\in L^2(\Gamma)$, we can make
out of it a state $\tilde\psi\in L^2(\r^n)$
by extending it constantly along the normals in $S_{t_0}$ and set it
equal to zero outside $S_{t_0}$, i.e.
$$\tilde\psi(\b x')=\psi\big(\alpha^{-1}\s\varphi(\b x').(\b x')\big)
\;\;\hbox{for}\;\;
\b x'\in S_{t_0}\;\;\hbox{and zero for }\;\;\b x'\not\in S_{t_0}$$
(cf. proof of lemma 3.2 for $\alpha$). 
Denote the space of these by $\cl H.^T_{t_0}$,
and the projection onto $\cl H.^T_{t_0}$ by $P^T_{t_0}$.
Then $\cl H.^T_{t_0}$ 
 is thought of as a ``thickening'' of $L^2(\Gamma)$
in $L^2(\r^n)$, and
note that $\cl H._{t_0}^T\cap C_\varphi$ is a transverse space.
However, when we deal with differential operators, the discontinuity 
at the boundary of the shell $S_{t_0}$ is a problem, so we prefer the following 
smooth version.
Let $\zeta_{t_0}\in C^\infty(\r)$
be a bump function which is one on $[-t_0,\,t_0]$ and zero outside
$[-t_0-\varepsilon,\,t_0+\varepsilon]$ for a given $\varepsilon$.
Then define for a $\psi\in L^2(\Gamma)$ the new thickening
$$\check\psi(\b x')=\zeta_{t_0}(\varphi(\b x'))\cdot\psi\left(
\alpha^{-1}\s\varphi({\bf x}).(\b x')\right)\;.$$
Denote the space spanned by these by $\cl H.^\zeta_{t_0}\subset L^2(\r^n)$,
then clearly $P_{t_0}\cl H.^\zeta_{t_0}=\cl H.^T_{t_0}$.

We now want to constrain the Hamiltonian in such a way that it can be
thought of as ``projecting the force down to its tangential component,''
where a Hamiltonian tangential to $\Gamma$ will be one which keeps
tangential motion to $\Gamma$ tangential, i.e. preserves the tangential
states on some shell $S_{t_0}$.
Recall that in the GNH--algorithm [GNH]
one restricted the Hamiltonian to all states with motion tangential
to $\Gamma$. To do the same here, we can now use either
of the two proposed methods $(\rn1)$ or $(\rn2)$ above
with the choice of
transverse space as 
$$\cl T._{t_0}:=\cl H._{t_0}^\zeta\cap C_\varphi\cap
{\rm Dom}\,H\;.$$
As long as $C_c^\infty(\r^n)\subset{\rm Dom}\,H$, we have that
$\kappa(\cl T._{t_0})$ is dense, in which case this is indeed a
transverse space. So assuming the latter, and that
$H\cl T._{t_0}\subset C_\varphi$, we define
$$H_{(\rn1)}\cdot\kappa(\psi):=\kappa(H\psi)\;\;\forall\,
\psi\in\cl T._{t_0}\,.\eqno{(4.1)}$$
In fact, when $\psi$ is smooth and $H$ is a differential operator, this becomes
$H_{(\rn1)}(\psi\rest\Gamma)=(H\psi)\rest\Gamma$, and then the
method is nothing but the one used by S. Helgason pp 251--252 [He]
for the restriction of a differential operator to a submanifold.
This is the first reasonable method to consider.

To motivate the use of method $(\rn2)$,
 consider how a physicist might object to
(4.1). Since $H$ need not preserve $\cl T._{t_0}$, it is possible
that $H\psi$ has momentum which is not tangential to $\Gamma$,
in which case it seems there is a force acting on the particle,
forcing it off $\Gamma$. To project this force out
of the total force acting on the particle the restriction procedure
above may not be appropriate. Instead, we intend to use the Hilbert space
projections to first project out of $H$ the parts which affect the 
momentum component normal to the level surfaces of $\varphi$
in the shell $\varphi^{-1}[-t_0,\, t_0]$,
then in the limit of $\kappa$ when the thickness of the shell goes to
zero we will be left with the appropriate Hamiltonian on $L^2(\Gamma)$
(having maintained tangentiality of momentum to $\Gamma$
during the limiting process).
In the light of this we define
$H_{(\rn2)}$ by:
$$\left(H_{(\rn2)}\cdot\kappa(\psi_1),\,\kappa(\psi_2)\right)\s L^2(\Gamma).
:=\lim_{t\to 0}{1\over 2t}\left(P^T_tH\psi_1,\;\hat hP\s t.
\psi_2\right)\eqno{(4.2)}$$
for all $\psi_1\in\cl T._{t_0}$ for which this exists
for all $\psi_2\in\cl T._{t_0}$, (so that
(4.2) defines a vector $H_{(\rn2)}\kappa(\psi_1)\in L^2(\Gamma)$).
Since $P_t\overline\cl T._{t_0}
\subset
\cl H._t^T$, (4.2) coincides with method $(\rn2)$.
This is the second reasonable method.

One can easily conceive of other methods apart from
$(\rn1)$ and $(\rn2)$ for constraining dynamics, for instance
forcing the particle onto $\Gamma$ by infinite potential walls:
\item{$(\rn3)$} Given a system $L^2(\r^n),\;\varphi,\; H$ as before,
restrict the particle to a box around $\Gamma$, say the shell
$S_{t}$, but for it to be well--defined
we need to assume $HC_c^\infty(S_t)\subseteq C_c^\infty(S_t)$
and that we have a selfadjoint extension $H_t$ on $L^2(S_t)$
of the symmetric operator $H\rest\set\psi\in C_c^\infty(S_t),
\psi(\varphi^{-1}(\pm t))=0.$. In the case when
$H$ is a differential operator, 
this means
one needs to decide how it behaves at the walls. 
Henceforth we assume $H$ is a differential operator. If we choose
ordinary reflection at the walls
(quantum billiards), then a $\psi\in{\rm Dom}\,H_t$
must satisfy
$${\partial\psi\over\partial\varphi}=0\qquad\hbox{when}\qquad\varphi=\pm t\;,$$
but without more detailed knowledge of $H$ we do not know what additional
boundary conditions $\psi$ must satisfy to make $H_t$ selfadjoint.
So ${\rm Dom}\, H_t$ is a subspace of
$$\cl D._t:=\set\psi\in P_tC^\infty(\r^n),{\psi\in L^2(\r^n),\;\;
\f\partial\psi,\partial\varphi.=0\;\;\hbox{when}\;\;\varphi(\b x')
=\pm t}.\subset C_\varphi.$$ Then in the limit when $t$ goes to zero, the boxes
force the particle (hence $H_t$) onto $\Gamma$. 
The appropriate way to take this
limit, is through $\kappa$, i.e. 
$$\left(H_{(\rn3)}\cdot\kappa(\psi_1),\,\kappa(\psi_2)\right)\s
L^2(\Gamma). := \lim_{t\to 0}{1\over 2t}\left(H_tP_t\psi_1,
\,\hat hP_t\psi_2\right)\eqno{(4.3)}$$
where $\psi_i\rest S_t\in{\rm Dom}\, H_t$ for all $t\in(0,\,t_0)$ and some
fixed $t_0$. Note that the possible $\psi_i\hbox{'s}$ are restricted by
the behaviour at the boundary. So here for reflection, we get that
$(\partial\psi_i/\partial\varphi)(\b y',\,t)=0$ for all $t\in[-t_0,\,0)\cup
(0,\,t_0]$. We recognise that $P_t\psi$ is in our transverse
space of earlier on.
When the space of $\kappa(\psi)$ for such $\psi$ is
dense in $L^2(\Gamma)$, $H_{(\rn3)}$ is
well--defined.
Other versions of this is possible if we change the behaviour at the
walls, e.g. introduce a phase with the reflection.

\noindent We will not discuss any other methods, but now the surprise is
that we have the following equivalences:

\thrm Theorem 4.4."$(1)$ With the assumptions above, 
$H_{(\rn1)}=H_{(\rn2)}=:H_\Gamma$,\chop
$(2)$ moreover if $H$ is a differential operator, then
$H_{(\rn1)}=H_{(\rn3)}$ on
${\rm Dom}\,H_{(\rn3)}$."
(1) We first show $H_{(\rn1)}=H_{(\rn2)}$.
Recall that $C_c^\infty(\r^n)\subset{\rm Dom}\,H$, and that
$\kappa(\cl T._{t_0}\cap C_c^\infty(\r^n))=C_c^\infty(\Gamma)$
hence it suffices to show that
$$\big((H_{(\rn1)}-H_{(\rn2)})\kappa(\psi_1),\,\kappa(\psi_2)\big)_{
L^2(\Gamma)}=0$$
for all $\psi_i\in \cl T._{t_0}\cap C_c^\infty(\r^n)$. By (4.1) and
(4.2), the left hand side is:
$$\lim_{t\to 0}{1\over 2t}\big((\un-P_t^T)H\psi_1,\,\hat hP_t\psi_2\big)
=\lim_{t\to 0}{1\over 2t}\int_{S_t}\overline{(\un-P_t^T)H\psi_1}\,
\psi_2\,\big|\nabla\varphi\big|\,d\mu$$
Partition $\Gamma$ into patches $U$ on which local coordinates exist
as in Lemma 3.2, then the right hand side becomes a finite sum of terms:
$$\lim_{t\to 0}{1\over 2t}\int_{-t}^td\varphi\,\int_Ud\b y'\,
\overline{(\un-P_t^T)H\psi_1}\,
\psi_2\;|\nabla\varphi|\,.\eqno{(*)}$$
(alternatively, take $\psi_2$ with support in $U$ and let $U$ vary)
Now $(\un-P_t^T)H\psi_1=:\rho_t\perp\cl H._t^T$, so for all
$\phi\in\cl H._t^T\cap C_c^\infty(S_t)$ we have
$$0=(\rho_t,\,\phi)=\int_Ud\b y'\,\phi(\b y')
\int_{-t}^td\varphi\,\overline\rho_t$$
where we used the fact that on $S_t$, $\phi$ is independent of $\varphi$,
so since this holds for all $\phi$ (which will span a dense subspace
of $L^2(U)$), we have that $\int_{-t}^t\rho_t(\b y',\,\varphi)\,d\varphi=0$
a.e. in $\b y'$. Now using the continuity of $\psi_2|\nabla\varphi|$, $(*)$
becomes:
$$\lim_{t\to 0}{1\over 2t}\int_Ud\b y'\,
\big(\psi_2|\nabla\varphi|\big)(\b y',\,
0)\,\int_{-t}^td\varphi\,\overline\rho_t=0\;.$$
Since this holds for all $U$, we find that $H_{(\rn1)}=H_{(\rn2)}$.\chop
(2) Next we prove that $H_{(\rn1)}=H_{(\rn3)}$ on ${\rm Dom}\, H_{(\rn3)}$.
Clearly $\bigcap\limits_{t\in[0,\, t_0]}\cl D._t
\rest S_{t_0}=\cl H._{t_0}^T$,
so it suffices to show that 
$$\leqalignno{\big((H_{(\rn1)}-H_{(\rn2)})\kappa(\psi_1),
\,\kappa(\psi_2)\big)_{L^2(\Gamma)}
&=0\qquad\forall\;\psi_i\in C_c^\infty(S_{t_0})\cap\bigcap_{t\in(0,t_0]}
{\rm Dom}\, H_t\cr
\lim_{t\to 0}{1\over 2t}\big((H&-H_tP_t)\psi_1,\,\hat hP_t\psi_2
\big)=0\;.&\hbox{i.e. that}\cr
\big((H-H_tP_t)\psi_1,\,\hat hP_t\psi_2\big)&=
\big((P_tH-H_tP_t)\psi_1,\,\hat hP_t\psi_2\big)\;,&\hbox{Now}\cr
\cl Q._t=:&\set\phi\in C_c^\infty(S_t),\phi(\varphi^{-1}(\pm t))=0.&
\hbox{and let}\cr
}$$
which is dense in $L^2(S_t)$, and $H=H_t$ on $\cl Q._t$.
Now given an open precompact set $X\subset S_t$, we can write any
$\phi\in C_c^\infty(\r^n)$ as $\phi=\phi_X+\widetilde\phi_X$
where $\phi_X\in\cl Q._t$ and $\phi\rest X=\widetilde\phi_X\rest X$.
Let $\psi\rest S_t\in{\rm Dom}\,H_t\cap C_c^\infty(S_t)$ 
for all $t\in(0,\,t_0]$, let $\phi\in\cl Q._t$ and $X$ be the interior
of ${\rm supp}\,\phi$. Then
$$\eqalignno{
\big((P_tH-H_tP_t)\psi,\,\phi\big)&=
\big((P_tH-H_tP_t)(\psi_X+\widetilde\psi_X),\,\phi\big)\cr
&=\big((H-H_t)\psi_X+(P_tH-H_tP_t)\widetilde\psi_X,\,\phi\big)\cr
&=\big((P_tH-H_tP_t)\widetilde\psi_X,\,\phi\big)&{(\dagger)}\cr}$$
Since $H$ and $H_t$ are differential operators, they preserve
supports, hence ${\rm supp}\big((P_tH-H_tP_t)\widetilde\psi_X\big)
\subseteq{\rm supp}\,\widetilde\psi_X$ which is disjoint from
the support of $\phi$, hence since the inner product in $(\dagger)$
is
an integral, we conclude that $(\dagger)$ is zero, i.e.
$(P_tH-H_tP_t)\psi\perp\cl Q._t$ hence ${(P_tH-H_tP_t)\psi} =0$.

We
 remark that any reasonable method of dynamics reduction should
produce a constrained Hamiltonian
$H_\Gamma$ on $L^2(\Gamma)$ which is essentially
selfadjoint and
the time evolution it generates,
$\exp it\overline{H}_\Gamma$, must preserve the algebra of observables 
$\cl R._S$ when we have obtained this (see next section).
We now test the above methods on two examples.
\itemitem{\bf Example 4.5.} Recall the previous example of the sphere
of radius $a$,
$\Gamma=\S^2$ in $\r^3$; $\cl H.=L^2(\r^3)=L^2(\r_+,\, r^2dr)\otimes
L^2(\Omega)$ with Hamiltonian 
$H=\hat\b p'^2/2m=-\f 1,2m.\Delta$, having domain of
essential selfadjointness $C_c^\infty(\r^3)$ and which acts on the
decomposable functions $\psi(\b x')=f(r)\cdot\xi(\theta,\,\phi)
\in C_c^\infty(\r^3)$ by
$$(-\Delta\psi)(\b x')=\left(-{d^2\over dr^2}-{2\over r}\cdot{d\over dr}
\right)f(r)\cdot\xi(\theta,\,\phi)+{1\over r^2}f(r)\cdot
B\xi(\theta,\,\phi)$$
where $B=|\b L'|^2$ is the Laplace--Beltrami operator on $L^2(\Omega)$
($\Omega$ denotes the unit sphere).
Now  for $t\in(0,\,a)$ we have
$$\cl H._t^T:=\set\psi\in\cl H.,{\psi(\b x')=\chi\s[\sqrt{a^2-t},\,
\sqrt{a^2+t}].(r)\cdot\xi(\theta,\,\phi)\;,\quad\xi\in L^2(\Gamma)}..$$
To obtain the smoothed space, choose a $\zeta_t\in C_c^\infty(\r)$
which is one on $[-t,\,t]$ and zero on $[-t-\varepsilon,\,t+\varepsilon]$
and set
$$\cl H._t^\zeta:=\set\psi\in\cl H.,{\psi(\b x')=\zeta_t(r^2-a^2)\cdot
\xi(\theta,\,\phi)\;,\quad\xi\in L^2(\Gamma)}.$$
and note that if $\psi\rest\Gamma=\xi\in C^\infty(\Gamma)$, then
$\psi\in C^\infty(\r^n)\subset\dom H$. So we choose the transverse
space $\cl T._t=\set\psi\in\cl H._t^\zeta,\psi\rest\Gamma\in C^\infty(\Gamma).$.
Now for method $(\rn1)$ for constraining $H$, let $\psi=(\zeta_t
\circ\varphi)\cdot
\xi\in\cl T._t$. Then from the explicit formula for $H$, we see
$$(H\psi)(\b x')={1\over 2mr^2}(B\xi)(\theta,\,\phi)\cdot\chi\s(\sqrt{a^2-t},
\,\sqrt{a^2+t}).(r) +\rho(\b x')$$
where $\rho$ is a function with support disjoint from
$\varphi^{-1}(-t,\, t)$. So, since $H\psi$ is continuous near $\Gamma$,
we have $\kappa(H\psi)=(H\psi)\rest\Gamma$, i.e.
$\kappa(H\psi)(\theta,\,\phi)=(B\xi)(\theta,\,\phi)/2ma^2$.
So $$H_{(\rn1)}\kappa(\psi)=\kappa(H\psi)={1\over 2ma^2}B\kappa(\psi)$$
for all $\psi\in\cl T._t$, so, since $\kappa(\cl T._t)=C^\infty(\Gamma)$
is dense in $L^2(\Gamma)$, $H_{(\rn1)}\xi 
=\f 1,2ma^2.B\xi$ for all $\xi\in
C^\infty(\Gamma)$ is densely defined, agrees with the classical
Hamiltonian $|\b L'|^2/2ma^2$ obtained in Sect. 2 (eq. (2.4)),
 and is essentially
selfadjoint. A good result.\chop
Next, as an exercise in method $(\rn2)$, we calculate
$$\left(P_t^TH\psi_1,\,\hat hP\s t.
\psi_2\right)\;,\quad\psi_i\in\cl T._{t_0},$$
where $\psi_i=(\zeta_t\circ\varphi)\cdot\xi_i$ as above. Now
$$(P_t^TH\psi_1)(\b x')= \left(P_t^T{B\over 2mr^2}\right)\xi(\theta,\,\phi)$$
if $r\in[\sqrt{a^2-t},\,\sqrt{a^2+t}]=:I_t$ and zero otherwise. To work out
the projection, we only need to consider the radial coordinate. The one
dimensional space $N$ generated by $\chi\s I_t.\in L^2(\r_+,\,r^2dr)$
corresponds to $\cl H._t^T$. So we need to decompose the function
$\chi\s I_t.(r)/2mr^2$ according to $N\oplus N^\perp$, i.e.
$\f 1,2mr^2.\chi\s I_t.(r)=\lambda_t\cdot\chi\s I_t.(r) + h(r)$
where $\lambda_t$ is a constant and $h\in N^\perp$, i.e.
$\int_{I_t}r^2h(r)\,dr=0$.
Now $(\chi\s I_t.,\,\f 1,2mr^2.\chi\s I_t.)=\f 1,2m.\int_{I_t}dr
=\lambda_t\int_{I_t}r^2dr$. i.e.
$$\eqalignno{\lambda_t&={3(\sqrt{a^2+t}-\sqrt{a^2-t})\over
2m[(a^2+t)^{3/2}-(a^2-t)^{3/2}]}&\hbox{and so:}\cr
(P_t^TH\psi_1)(\b x')&=\lambda_t\cdot\chi\s I_t.(r)\cdot(B\xi_1)
(\theta,\,\phi)\,.\cr
\hbox{Thus}\qquad\quad\big(P_t^TH\psi_1,&\,
\hat hP\s[-t,\,t].\psi_2\big)\cr
&=\int\lambda_t\cdot\chi\s I_t.(r)(B\overline\xi_1)(\theta,\,\phi)\cdot
2r\psi_2(\b x')r^2dr\,d\Omega\cr
&=2\lambda_t\int_{\sqrt{a^2-t}}^{\sqrt{a^2+t}}r^3dr\int(B\overline\xi_1)\cdot
\xi_2\,d\Omega\cr
&=2\lambda_t{1\over 4a^2}\big((a^2+t)^2-(a^2-t)^2\big)\cdot
(B\xi_1,\,\xi_2)\s L^2(\Gamma).\cr
&=2t\lambda_t\cdot(B\xi_1,\,\xi_2)\s L^2(\Gamma).\cr}$$
where we used the fact that the measure on $\Gamma$ is $a^2d\Omega$
with $d\Omega$ the usual measure on the unit sphere.
Now
$$\leqalignno{\lim_{t\to 0}\lambda_t&={3\over 2m}\lim_{t\to 0}
{\sqrt{a^2+t}-\sqrt{a^2-t}\over(a^2+t)^{3/2}-(a^2-t)^{3/2}}
={1\over2ma^2}\,,\quad\hbox{hence}\cr
\lim_{t\to 0}{1\over 2t}\big(P_t^TH\psi_1&,\,\hat hP\s[-t,\,t].\psi_2\big)
=\lim_{t\to 0}\lambda_t\big(B\xi_1,\,\xi_2\big)\s L^2(\Gamma).\cr
&=\left({B\over 2ma^2}\xi_1,\,\xi_2\right)\s L^2(\Gamma).
=(H_{(\rn2)}\xi_1,\,\xi_2)\qquad
\hbox{for all $\xi_i\in C^\infty(\Gamma)$.}\cr}$$
Thus $H_{(\rn2)}=B/2ma^2=H_{(\rn1)}=H_\Gamma$ on $C^\infty(\Gamma)$, 
in agreement with theorem 4.4.

\noindent 
Without much extra effort,
we can also show that Schr\"odinger operators
$H={1\over 2m}\hat\b p'^2+V(\hat\b q')$ for $V$ smooth near $\Gamma$
produce for $\xi\in C^\infty(\Gamma)$:
$$H_{(\rn1)}\xi(\theta,\,\phi)=H_{(\rn2)}\xi(\theta,\,\phi)
=\left({B\over 2ma^2}\xi\right)(\theta,\,\phi)+V(a,\,\theta,\,\phi)\cdot
\xi(\theta,\,\phi).$$

\itemitem{\bf Example 4.6.} We would like to restrict a free quantum
particle in $\r^3$ to a cylinder $\Gamma$
 of radius $a$ around the $z\hbox{--axis}$.
We do this in exact analogy with the sphere.
Let the constraint be $\varphi(\b x')=x^2+y^2-a^2$, then
$\nabla\varphi(\b x')=(2x,\,2y,\,0)$ hence the critical points
are well away from $\Gamma$. So
$$\varphi^{-1}[-s,\,t]=\set\b x'\in\r^3,r^2\in[a^2-s,\,a^2+t].$$
where we henceforth use cylindrical coordinates $(z,\,r,\,\theta)$.
Then
$$\eqalignno{(\psi_1,\,\psi_2)_\Gamma&=
\lim_{t\to 0}{1\over 2t}\big(\psi_1,\,\hat hP_t\psi_2\big)
\cr
&=\hlf{d\over dt}\int^{\sqrt{a^2+t}}_{\sqrt{a^2-t}}
2r^2\,dr\int_{-\infty}^\infty dz\int_0^{2\pi}d\theta(\overline\psi_1
\psi_2)(\b x')\Big|_{t=0}.\cr}$$
Let $\psi_i$ be decomposable $\psi_i(\b x')=f_i(r)\xi_i(z,\,\theta)$
where $\xi_i\in L^2(\Gamma)$ and $f_i$ is continuous at $r=a$.
Then 
$$\eqalignno{(\psi_1,\,\psi_2)_\Gamma&=
\int\overline\xi_1\xi_2\,dz\,d\theta\cdot
{d\over dt}\int^{\sqrt{a^2+t}}_{\sqrt{a^2-t}}r^2\overline
f_1f_2(r)\,dr\Big|_{t=0}\cr
&=a\overline f_i(a)f_2(a)\int\overline\xi_1\xi_2\,d\theta\,dz\,.\cr}$$
The free Hamiltonian $H=-\Delta/2m$ in cylindrical coordinates acts by:
$$\Delta\psi={1\over r}{\partial\over \partial r}\left(r{\partial\psi
\over\partial r}\right)+{1\over r^2}{\partial^2\psi\over\partial\theta^2}
+{\partial^2\psi\over\partial z^2}\,.$$
For $t\in(0,\,a)$ we have:
$$\cl H._t^T=\set\psi\in\cl H.,\psi(\b x')={\chi\s I_t.}(r)\cdot
\xi(z,\,\theta)\,,\;\;\xi\in L^2(\Gamma).$$
where $I_t:=[\sqrt{a^2-t},\,\sqrt{a^2+t}]$. Then with a bump function
$\zeta_t\in C^\infty(\r)$ which is one on $[-t,\,t]$ and zero outside
$[-t-\varepsilon,\,t+\varepsilon]$, we have
$$\cl H._t^\zeta:=\set\psi\in\cl H.,\psi(\b x')=\zeta_t(r^2-a^2)\cdot
\xi(z,\,\theta)\,,\;\;\xi\in L^2(\Gamma).\,.$$
Since the smooth functions of compact support are in the domain of $H$,
we choose
$$\cl T._t=\set\psi\in\cl H._t^\zeta,
\psi\rest\Gamma\in C_c^\infty(\Gamma).\,.$$
Now to obtain $H_{(\rn1)}$, let $\psi=(\zeta_t\circ\varphi)\cdot\xi
\in\cl T.$, then
$$(H\psi)(\b x')=\left(\f -1,2m.\right)\chi\s J_t.(r)\left(
{1\over r^2}\,{\partial^2\over\partial\theta^2}+{\partial^2\over
\partial z^2}\right)\xi(z,\,\theta)
+\rho(\b x')$$
where $J_t:=(\sqrt{a^2-t},\,\sqrt{a^2+t})$ and $\rho$ has support disjoint from
$J_t$. Now since $\kappa(H\psi)=(H\psi)\rest\Gamma$, we get
$$H_{(\rn1)}\xi(z,\,\theta)=
\left(\f -1,2m.\right)\left(
{1\over a^2}\,{\partial^2\over\partial\theta^2}+{\partial^2\over
\partial z^2}\right)\xi(z,\,\theta)$$
for all $\xi\in C_c^\infty(\Gamma)$. 
Thus via theorem 4.4 $$H_\Gamma=
\left({-1\over 2m}\right)\left(
{1\over a^2}\,{\partial^2\over\partial\theta^2}+{\partial^2\over
\partial z^2}\right)$$
on $C_c^\infty(\Gamma)$. 
This is precisely the quantization one would expect
of the constrained classical Hamiltonian (2.6).

\item{\bf Remarks.}(1) We regard the choice of a transverse space 
$\cl T.$ as a decision on the direction from which $H$ should be
reduced to $\Gamma$. Above we chose the normal direction.
In quantum systems with less geometry (e.g. no metric on an 
underlying manifold), it may be difficult to decide on an
appropriate $\cl T.$. This choice has physical content, and seems
analogous to the choice of a selfadjoint extension for a
Hamiltonian in the Dirac approach sketched early this section.
An alternative way of expressing the choice of the transverse
space is to observe that on $L^2(S_{t_0})$ for $t_0$ small enough,
the pair of operators $\hat\varphi$ and $P_\varphi:=i\partial/\partial\varphi$
form a canonical pair, i.e. $[P_\varphi,\,\hat\varphi]\psi=
i\psi$ for all smooth $\psi$ with compact support in the interior
of $S_{t_0}$. Then $\cl T.\rest S_{t_0}=\ker P_\varphi\cap
\dom(H\rest L^2(S_{t_0}))$. So in a general quantum system one
can look for a given ``local canonical conjugate'' to the
constraint to obtain a transverse space as its kernel.
In fact, this observation also gives a clue on how to enforce
second--class constraints, in the sense that if we are given a
canonical pair $\varphi,\; P_\varphi$ to enforce, then we do it
as above, by using $P_\varphi$ to {\it locally} select a transverse
space to $\varphi^{-1}(0)$.
\item{(2)} Recall that in Sect. 2, the classical secondary 
constraint we obtained for $\varphi$ was $(\nabla\varphi)\cdot
\b p'\rest\Gamma$. The selection of the normal transverse space 
above, can be thought of as the enforcement of the constraint
$\nabla\varphi\cdot\hat\b p'\psi=0$ near $\Gamma$, which looks
very much like what one would expect a ``quantization'' of the
classical secondary constraint to be. This leads one to ask whether
we can quantize the infinitesimal Dirac procedure which produced
this secondary constraint as a whole. Unfortunately this does not
seem to work;-- we sketch how it goes awry.
Given the primary constraint $\hat\varphi$ and the Hamiltonian
$H=\f 1,2m.\hat\b p'^2+V(\hat\b q')$, the secondary constraint
should be $[H,\,\hat\varphi]$ ``restricted'' to $\Gamma$.
Now $$[H,\,\hat\varphi]\psi=-\f 1,2m.(\Delta\varphi)\psi
-\f 1,m.(\nabla\varphi)\cdot(\nabla\psi)$$
for all $\psi\in C_c^\infty(\r^n)$. 
Restriction to $\Gamma$ is done by $\kappa$.
However, whilst $-\f 1,m.(\nabla\varphi)\cdot(\nabla\psi)$
is proportional to the desired secondary constraint
$\nabla\varphi\cdot\hat\b p'\psi$, the 
term $-\f 1,2m.(\Delta\varphi)\psi$ need not vanish on $\Gamma$.
Moreover if we use $[H,\,\hat\varphi]$ as a new secondary
constraint to select the transverse space instead of
$\nabla\varphi\cdot\hat\b p'$, we obtain the wrong result
on the sphere.
Nevertheless, given the close parallel which the selection of
the transverse space has to the classical secondary constraint,
we regard the use of a transverse space here as the imposition
of a secondary quantum constraint.
\item{(3)} We remark that one needs to reduce the
dynamics infinitesimally, i.e. through the Hamiltonian,
not directly on the time evolution unitaries
$\exp(itH)$.
\item{(4)} In general, there is no guarantee that the constraining
of a selfadjoint Hamiltonian $H$ to $\Gamma$ produces an
essentially selfadjoint operator on $L^2(\Gamma)$.
This is a difficult question that needs further investigation,
and its classical equivalent in the GNH--algorithm also
is unsolved, that is, we do not know whether reducing
a given complete Hamiltonian vector field to a submanifold
$\Gamma$ produces a complete vector field on $\Gamma$.
When $\Gamma$ has edges, it is very easy to get
examples of complete vector fields which will not be complete
when constrained to $\Gamma$.
\item{(5)} The confining potential approach, developed in
[Ma, Fu], also treats the dynamics of quantum particles on
(or near) surfaces. In this, one assumes there is a confining 
potential depending only on the normal coordinates, and with a
deep minimum on $\Gamma$. The system is then taken to be in a fixed
eigenstate in the normal direction (``normal degrees of freedom
are frozen''), and this was applied to some physical examples where
such confining potentials actually occur [Ma].
By contrast, our object is to 
 impose the constraints
exactly in order to understand the nature of the mathematical
problems involved with the construction of a quantum version of
Dirac's theory of constraints. As can be seen from this paper, 
such a program encounters deep pathologies which do not arise when
the potentials only provide approximate confinement.
Furthermore if one has a physical principle 
which requires the particle to be exactly
on $\Gamma$, or if one is defining quantum mechanics on a curved manifold,
it also seems more appropriate to solve the constraint exactly. 
In addition to the dynamics, we also do the kinematics.

\beginsection 5. Constrained Observables.

In this section we continue the analysis 
of the problem of the previous sections
specifically in regard to the observables. That is, given the unbounded
map $\kappa:C_\varphi\to L^2(\Gamma)$ above, we wish to examine how
selfadjoint operators and unitaries on $L^2(\r^n)$ lift through $\kappa$
to produce operators on $L^2(\Gamma)$. Due to the unboundedness 
and nonclosability of $\kappa$
there will be some pathology, even for bounded operators on $L^2(\r^n)$.

The choice of field algebra $\cl F.\subseteq\cl B.(\cl H.)$
will turn out to be important for obtaining a nontrivial constrained
field algebra on $L^2(\Gamma)$. In fact, the CCR--algebra
$\overline{\Delta(\r^{2n})}=C^*\set\exp(i\hat{\b q'}\cdot\b a'){\,,\;}
\exp(i\hat{\b p'}\cdot\b a'),\b a'\in\r^n.$
will be too small if $\Gamma$ is a curved manifold
(see later in this section). A more suitable field algebra for curved
$\Gamma$ is $C_b(\r^n)\rtimes{\rm Diff}\,\r^n$, which is here concretely
the C*--algebra generated in $\cl B.(L^2(\r^n))$ by the multiplication
operators $\set T_f,f\in C_b(\r^n).$, 
$$(T_f\psi)(\b x'):=f(\b x')\psi(\b x')\;,\qquad\psi\in L^2(\r^n)=\cl H.$$
and the set of unitaries
$\set V_\beta,\beta\in{\rm Diff}\,\r^n.$, where
$$(V_\beta\psi)(\b x'):=J_\beta(\b x')^{1/2}\psi(\beta\b x')\,,\quad\psi
\in L^2(\r^n)$$
with $J_\beta$ the Jacobian of $\beta\in{\rm Diff}\,\r^n$.
We will concentrate on these two classes of operators below.

Now an operator $A\in\cl B.(\cl H.)$ will lift through $\kappa$
to a densely defined operator $\Lambda(A)$ on $L^2(\Gamma)$ if:
\item{$\bullet$} there is a space $S\subset C_\varphi$ such that
$AS\subseteq C_\varphi$ and $\kappa(S)$ is dense in $L^2(\Gamma)$,
\item{$\bullet$} $A\psi\in\ker(\cdot,\cdot)_\Gamma=\ker\kappa$ for all
$\psi\in S\cap\ker\kappa$.

\noindent In this case $$\Lambda(A)\kappa(\psi):=\kappa(A\psi)\;\;\forall\,
\psi\in S,\eqno{(5.1)}$$
or equivalently
$$ \big(\Lambda(A)\kappa(\phi),\,\kappa(\psi)\big)_\Gamma=
\lim_{t\to 0}{1\over 2t}(A\phi,\,\hat hP_t\psi)\;\;
\forall\,\psi,\,\phi\in S.$$
Since we are interested in *--algebras of operators, we will concentrate on
situations where the dense subspace $S$ is invariant under the class of
operators under consideration. Define the *--algebra:
$$\cl F._S:=\set A\in\cl F.,A\psi\in S\ni A^*\psi\;\;\forall\,\psi\in S.\;.
\eqno{(5.2)}$$
There are three obvious dense subspaces $S$ we can ask to be preserved,
$C_\varphi$, $C_c(\r^n)$ and the transverse space $\cl T._t=\cl H._t^\zeta
\cap C_\varphi\cap{\rm dom}\, H$ of the last section
(or better still, the space $\cl T.^{(t_0)}$ 
spanned by all $\cl T._t$ for $t\in(0,\,t_0]$ 
and all possible smoothings $\zeta$),
 and we will consider all these in due course.
A useful way of characterising $\cl T.^{(t_0)}$ is as
$$\cl T.^{(t_0)}=\set\psi\in C_c(\r^n),{\partial\psi\over\partial\varphi}(\b x')
=0\quad\forall\;\b x'\in S_{t_\psi},\;\;0<t_\psi<t_0.\bigcap{\rm Dom}\, H\,.$$
The choices $\cl T._t$ or $\cl T.^{(t_0)}$ can be thought of as enforcing 
secondary quantum constraints on the observables, cf. remark 2 of Sect. 4.
However, we will see that in the quantum picture there is no compelling
reason to do this.

Unless we specify what $S$ is below, we will assume some choice has been made.
The elements of $\cl F._S$ which will lift via (5.1) are those which
preserve $S\cap\ker\kappa$, and for these we have:
$\Lambda(A)\Lambda(B)\kappa(\psi)=\kappa(AB\psi)=\Lambda(AB)\kappa(\psi)$
for all $\psi\in S$, i.e. $\Lambda$ is an algebra homomorphism.
However, there are two pathologies associated with $\Lambda$;--
\item{$(\rn1)$} given an $A\in\cl B.(\cl H.)$ for which $\Lambda(A)$
exists, then $\Lambda(A)$ need not be bounded,
\item{$(\rn2)$} given an $A\in\cl B.(\cl H.)$ for which both
$\Lambda(A)$ and $\Lambda(A^*)$ exist, we need not have that
$\Lambda(A^*)\subseteq\Lambda(A)^*$.

\noindent\b 5.3. Example of $(\rn1)$:' \chop
Let $\cl H.=L^2(\r^2)$, $\varphi(\b x')=x_2$ so $\Gamma=\varphi^{-1}(0)$ is
the $x_1\hbox{--axis.}$ Choose $S=C_c(\r^n)$.
Clearly all $V_\beta$ preserve $S=C_c(\r^2)$
for all $\beta\in{\rm Diff}\,\r^2$, so since $
(V_\beta^*\psi)(\b x')=J_\beta^{-1/2}(\beta^{-1}\b x')\cdot
\psi(\beta^{-1}\b x')$ 
we have $V_\beta\in\cl F._S$. If $\beta\Gamma\subseteq\Gamma$, we 
get that $V_\beta$ preserves $C_c(\r^2\backslash\Gamma)=\ker(\cdot,\cdot)_\Gamma
\cap C_c(\r^2)$, and so $\Lambda(V_\beta)$ exists.
Consider now the $\beta\in{\rm Diff}\,\r^2$ given by $\beta(x_1,\, x_2):=
(ax_1,\, x_2e^{x_1})$, $a>0$, so clearly $\beta\Gamma=\Gamma$
and $J_\beta(\b x')=ae^{x_1}\neq 0$.
(That $\beta$ is a diffeomorphism is clear since its inverse is 
$\beta^{-1}(x_1,\,x_2)=(x_1/a,\; x_2e^{-x_1/a})$ which is also differentiable).
Now for $\psi\in C_c(\r^2)$ we have
$$\eqalignno{\left(\Lambda(V_\beta)\kappa(\psi),\,\Lambda(V_\beta)\kappa(\psi)
\right)_{L^2(\Gamma)}&=(V_\beta\psi,\,V_\beta\psi)_\Gamma=
\left(V_\beta\psi\rest\Gamma,\,V_\beta\psi\rest\Gamma\right)_{L^2(\Gamma)}\cr
&=\int\big|(V_\beta\psi\rest\Gamma)\big|^2dx_1\cr
&=\int J_\beta(x_1,\, 0)\big|\psi(ax_1,\, 0)\big|^2dx_1\cr
&=\int e^{x/a}|\psi(x,\, 0)|^2 dx=:I_\psi\cr}$$
and now we can choose a $\psi\in C_c(\r^2)$ with 
$\|\kappa(\psi)\|=\|\psi\rest\Gamma\|\s{L^2(\Gamma)}.=1$
which can make $I_\psi$ arbitrary large, e.g. 
$\psi=\chi\s[n,\,n+1]\times[0,\,1].$ and consider
$I_\psi$ as $n\to\infty$. Thus $\Lambda(V_\beta)$ is
unbounded.

\medskip
We remark that if we chose a more restrictive space $S$, e.g.
$\cl T.^{(t_0)}$, then the diffeomorphism in the last example
will preserve $\cl T.^{(t_0)}$, so this pathology cannot be removed by
enforcing ``secondary quantum constraints.''
\def\L#1'{{\Lambda\big(#1\big)}}\chop
\noindent\b Example of $(\rn2)$:'\chop
Continue the previous example, noting that $\Lambda(V_\beta^*)$ exists
because $\beta^{-1}\Gamma=\Gamma$. However, if $\Lambda(V_\beta^*)
\subseteq\Lambda(V_\beta)^*$ we have for all $\psi\in C_c(\r^2)$ that
$$\eqalignno{\left(\L V_\beta'\kappa(\psi),\,\L V_\beta'\kappa(\psi)\right)\s
L^2(\Gamma).&=\left(\kappa(\psi),\,\L V_\beta'^*\kappa(V_\beta\psi)\right)\s
L^2(\Gamma).\cr
&=\left(\kappa(\psi),\,\L V_\beta^*'\kappa(V_\beta\psi)\right)\s L^2(\Gamma).\cr
&=\left(\kappa(\psi),\,\kappa(V_\beta^*V_\beta\psi)\right)\s L^2(\Gamma).\cr
&=\left(\kappa(\psi),\,\kappa(\psi)\right)\s L^2(\Gamma).\;,\cr}$$
which makes $\L V_\beta'$ unitary, hence bounded, in contradiction with
example $(\rn1)$, so it is false that $\Lambda(V^*_\beta)\subseteq
\Lambda(V_\beta)^*$. 

\medskip
Now, it appears reasonable to the authors that in a constraining method,
boundedness and the adjoint operation should be preserved, at least on
the physical variables.  So we want to restrict the set of operators
under consideration to those satisfying these two requirements.
Observe that an $A\in\cl F._S$ will lift to a bounded operator 
$\L A'$ on $L^2(\Gamma)$ iff
$$(A\psi,\, A\psi)_\Gamma\leq M\cdot(\psi,\,\psi)_\Gamma\;\;\forall\,
\psi\in S\quad\hbox{and a fixed}\quad M<\infty\,.$$
(Clearly if $\psi\in S\cap\ker\kappa$ then the last 
inequality implies $(A\psi,\, A\psi)_\Gamma=0$, i.e. $A\psi\in
\ker\kappa$, so $\L A'$ exists).
\chop\b Definition:'
$$\leqalignno{\cl O._S&:=\big\{A\in\cl F._S\,\Big|\;(A\psi,\,A\psi)_\Gamma\leq
M_A\cdot(\psi,\,\psi)_\Gamma\geq(A^*\psi,\,A^*\psi)_\Gamma
\cr
&\qquad\forall\,\psi\in S\quad\hbox{and some}\quad M_A<\infty\,,\quad
\hbox{and}\quad\L A^*'\subseteq\L A'^*\;\big\}\cr
\cl D._S&:=\set A\in\cl O._S,AS\subseteq\ker\kappa
\supseteq A^*S.\;.\cr}$$
Note that if $A\in\cl O._S$, then $\L A'$ is bounded, so we have in fact
that $\L A'^*=\overline{\L A^*'}$, so there are no problems with selfadjointness
if $A$ is selfadjoint.
These will occur however if we consider $\Lambda$ on unbounded operators.
\thrm Lemma 5.4." $\cl O._S$ and $\cl D._S$ are *--algebras."
Both sets are linear spaces. To see that $\cl O._S$ is closed under
taking of adjoints, take the adjoint of $\L A^*'\subseteq
\L A'^*$ to get $\L A^{**}'=\L A'\subseteq\overline{\L A'}=\L A'^{**}
\subseteq\L A^*'^*$, i.e. $A^*$ is in $\cl O._S$ if $A$ is.
Then clearly $\cl D._S$ is also closed under taking of adjoints.
To see that $\cl O._S$ is an algebra, let $A,\; B\in \cl O._S$, then
$$(AB\psi,\, AB\psi)_\Gamma\leq M_A\cdot(B\psi,\, B\psi)_\Gamma
\leq M_AM_B(\psi,\,\psi)_\Gamma$$
for all $\psi\in S$, and similarly for $B^*A^*$. Moreover
$$\L AB'^*=[\L A'\L B']^*\supseteq\L B'^*\L A'^*\supseteq
\L B^*'\L A^*'=\L(AB)^*'$$
and thus $AB\in\cl O._S$. That $\cl D._S$ is an algebra is obvious.

Now $\cl D._S=\ker(\Lambda\rest{\cl O._S})$ is a $*\hbox{--ideal}$
of $\cl O._S$, and $\cl R._S:=\cl O._S\big/\cl D._S\cong\L\cl O._S'\subset
\cl B.(L^2(\Gamma))$. We think of $\cl R._S$ as the ``physical observables''
obtained from enforcing the constraint on $\cl O._S$, in analogy to the
algebra $\cl R.$ of the T--procedure for Dirac constraining.
$\cl R._S$ cannot be zero because the identity operator
$\un\in\cl O._S$ and $\overline{\L\un'}=\un$.
We do not expect $\cl R._S$ to be a C*--algebra, but it can easily
generate a C*--algebra since $\cl R._S\cong\L\cl O._S'\subset\cl B.
(L^2(\Gamma))$.

Given that the commutant $\hat\varphi'$ 
is the traditional observables, we find here:
\thrm Theorem 5.5." $\cl F._S\cap\hat\varphi'\cap\hat h'\subset\cl O._S$."
Let $A\in\cl F._S\cap\hat\varphi'\cap\hat h'$, 
which is a *--algebra, so it also
contains $A^*$. Then
$$\eqalignno{(A\psi,\,\hat hP_tA\psi)&=\|\widehat{h^{1/2}}P_tA\psi\|^2
=\|A\widehat{h^{1/2}}P_t\psi\|^2\cr
&\leq\|A\|^2\cdot\|\widehat{h^{1/2}}P_t\psi\|^2=
\|A\|^2\cdot(\psi,\,\hat hP_t\psi)
\cr}$$
for all $\psi\in S$. Thus
$$\eqalignno{(A\psi,\,A\psi)_\Gamma&=\lim_{t\to 0}\f 1,2t.(A\psi,\,
\hat hP_tA\psi)\cr
&\leq\|A\|^2\lim_{t\to 0}\f 1,2t.(\psi,\,\hat hP_t\psi)
=\|A\|^2\cdot(\psi,\,\psi)_\Gamma\;.\cr}$$
The same is true for $A^*$. Now for all $\psi,\;\phi\in S$:
$$\eqalignno{\left(\L A'\kappa(\psi),\,\kappa(\phi)\right)\s L^2(\Gamma).
&=\left(\kappa(A\psi),\,\kappa(\phi)\right)\s L^2(\Gamma).\cr
=(A\psi,\,\phi)_\Gamma&=\lim_{t\to 0}\f 1,2t.(A\psi,\,\hat hP_t\phi)
\cr
&=\lim_{t\to 0}\f 1,2t.(\psi,\, A^*\hat hP_t\phi)
=\lim_{t\to 0}\f 1,2t.(\psi,\,\hat hP_tA^*\phi)\cr
&=(\psi,\, A^*\phi)_\Gamma =\big(\kappa(\psi),\,\L A^*'\kappa(\phi)
\big)\s L^2(\Gamma).\cr}$$
so $\L A^*'\subseteq\L A'^*$.
Thus $A\in\cl O._S$.

\item{\b Remarks.'}\b (1)' Note that due to the limit in $(\cdot,\cdot)_\Gamma$,
it is only the behaviour near $\Gamma$ which contributes in the constraining of
an operator. In fact the proof actually shows that an $A\in\cl F._S$
will be in $\cl O._S$ if 
$$\eqalignno{\lim_{t\to 0}\f 1,t.\|\widehat{h^{1/2}}P_{t}A\psi\|^2
&=\lim_{t\to 0}\f 1,t.\|A\widehat{h^{1/2}}P_{t}\psi\|^2\cr
\hbox{and}\qquad\qquad\lim_{t\to 0}\f 1,t.(\psi,\,[A^*,\,&\hat hP_{
t}]\phi
)=0\cr}$$
for all $\psi,\;\phi\in S$ and the same for $A^*$.
So if $A$ commutes with $\hat\varphi$ and $\hat{h}$ on a small
neighbourhood of $\Gamma$, it will also be in $\cl O._S$.
Behaviour of $A$ away from $\Gamma$ is irrelevant.
\item{\b (2)'} We note by 5.5  that all multiplication operators by bounded
Borel functions which preserve $S$ will be in $\cl O._S$.
In particular, for a reasonable physical system we expect $\hat\varphi$
to preserve $S$, so the constraint $\hat\varphi$ is in $\cl O._S$. 
In this case $\hat\varphi$ is in $\cl D._S$ by the following argument:
$$\eqalignno{\lim_{t\to 0}\|\widehat{h^{1/2}}P_{t}\hat\varphi\psi\|^2&=
\lim_{t\to 0}\f 1,t.\|\hat\varphi\widehat{h^{1/2}} P_{t}\psi\|^2\cr
&\leq\lim_{t\to 0}\f 1,t.\sup\set|\varphi(\b x')|^2,\b x'\in
\varphi^{-1}[-t,\,t].\cdot\|\widehat{h^{1/2}}P_t\psi\|^2\cr
&=\left(\lim_{t\to 0}\sup\set|\varphi(\b x')|^2,\b x'\in\varphi^{-1}[-t,\, t].
\right)\cdot \lim_{t\to 0}\f 1, t.\|\widehat{h^{1/2}}P_t\psi\|^2\cr
&=0\qquad\forall\,\psi\in S\cr}$$
using the facts that the limit $\lim\limits_{t\to 0}\f 1,t.
\|\widehat{h^{1/2}}
P_t\psi\|^2 =\lim\limits_{t\to 0}\f 1,t.(\psi,\,\hat hP_t\psi)$ exists due to
$\psi\in S\subset C_\varphi$, and that the limit of the supremum
is zero. Hence
$(\hat\varphi\psi,\,\hat\varphi\psi)_\Gamma=0$, i.e. $\hat\varphi\psi
\in\ker\kappa$ for all $\psi\in S$, so $\Lambda(\hat\varphi)=0$ as expected.

\thrm Theorem 5.6." An $A\in\cl F._S$ preserves $\ker\kappa\cap S$
iff 
$$\lim_{t\to 0}\f 1,t.\|P_{t}A(\un-P_{t})\psi\|^2=0$$
for all $\psi\in\ker\kappa\cap S$."
We need the following lemma:\chop
{\bf Lemma:} {\sl With assumptions and notation as above, we have\chop
$\psi\in\ker(\cdot,\cdot)_\Gamma$ 
iff $\lim\limits_{t\to 0}\f 1,t.\|P_t\psi\|^2
=0$.}\chop {\it Proof:}
$\psi\in\ker(\cdot,\cdot)_\Gamma$ 
iff $0=(\psi,\,\psi)_\Gamma=\lim\limits_{t\to 0}
\f 1,2t.(\psi,\,\hat hP_t\psi)$.
Now
$$\eqalignno{(\psi,\,\hat hP_t\psi)&=\int_{\varphi^{-1}[-t,t]}|\nabla
\varphi(\b x')|\cdot|\psi(\b x')|^2d\mu(\b x')\cr
&\geq\inf\set|\nabla\varphi(\b x')|,\b x'\in\varphi^{-1}[-t,\, t].\,
\int_{\varphi^{-1}[-t,t]}|\psi(\b x')|^2d\mu(\b x')\;.\cr}$$
Since $|\nabla\varphi(\b x')|$ is assumed to be bounded, continuous and nonzero
on a neighbourhood of $\Gamma$ we have
$$\leqalignno{\lim_{t\to 0}\inf
\set|\nabla\varphi(\b x')|,\b x'\in\varphi^{-1}[-t,\, t].
&=:M>0\,.\cr
\lim_{t\to 0}{1\over t}(\psi,\,\hat hP_t\psi)\geq
M\lim_{t\to 0}{1\over t}\int_{\varphi^{-1}[-t,t]}&|\psi(\b x')|^2d\mu(\b x')
&\hbox{Thus}\cr
=M\lim_{t\to 0}{1\over t}&\|P_t\psi\|^2\cr
\hbox{so}\quad\psi\in\ker(\cdot,\cdot)_\Gamma\quad\hbox{implies}\quad
0=\lim_{t\to 0}{1\over t}(\psi,\,\hat hP_t\psi)\geq M\lim_{t\to 0}&
{1\over t}\|P_t\psi\|^2\cr
\hbox{and so:}\qquad\lim_{t\to 0}{1\over t}\|P_t\psi\|^2&=0\cr}$$
On the other hand, given a $\psi\in\cl H.$ satisfying the last 
equation, we have
$$\eqalignno{(\psi,\,\hat hP_t\psi)&=\int_{\varphi^{-1}[-t,t]}
|\nabla\varphi(\b x')|\cdot|\psi(\b x')|^2d\mu(\b x')\cr
&\leq\sup\set |\nabla\varphi(\b x')|,\b x'\in\varphi^{-1}[-t,\,t].
\int_{\varphi^{-1}[-t,t]}|\psi(\b x')|^2d\mu(\b x')\cr}$$
and the the limit of the
 supremum (denoted $N$) exists by the boundedness assumption,
and is strictly positive. So
$$\lim_{t\to 0}{1\over t}(\psi,\,\hat hP_t\psi)\leq N\cdot\lim_{t\to 0}
{1\over t}\|P_t\psi\|^2=0$$
i.e. $\lim\limits_{t\to 0}{1\over t}(\psi,\,\hat hP_t\psi)=0$,
i.e. $\psi\in\ker(\cdot,\cdot)_\Gamma$.
\hfill$\blacktriangledown$\break
\chop
So by this lemma, $A\in\cl F._S$ preserves $\ker\kappa\cap S$
iff $\lim\limits_{t\to 0}\f 1,t.\|P_{t}A\psi\|^2=0$ for all
$\psi\in\ker\kappa\cap S$. Now
$$\eqalignno{\lim_{t\to 0}\f 1,t.\|P_tAP_t\psi\|^2&\leq
\lim_{t\to 0}\f 1,t.\|P_tAP_t\|^2\|P_t\psi\|^2\cr
&\leq\|A\|^2\cdot\lim_{t\to 0}\f 1,t.\|P_t\psi\|^2=0\cr}$$
whenever $\psi\in\ker\kappa$. So by the triangle inequality:
$$\leqalignno{\|P_tAP_t\psi\|+\|P_tA(\un-P_t)\psi\|&\geq\|P_tA\psi\|\cr
&\geq\big|\|P_tAP_t\psi\|-\|P_tA(\un-P_t)\psi\|\big|\cr
\lim_{t\to 0}\f 1,\sqrt{t}.\|P_tA\psi\|&=\lim_{t\to 0}
\f 1,\sqrt{t}.\|P_tA(\un-P_t)\psi\|&\hbox{we obtain:}\cr}$$
for all $\psi\in\ker\kappa\cap S$.

An immediate consequence is that $\Lambda(A)$ exists for all
$A\in\hat\varphi'\cap\cl F._S$ (though $\Lambda(A)$ need not be
bounded in general), and as before, we only need
commutativity close to $\Gamma$ to get this.
Also note the similarity with the condition ${P_{\rm phys}A(\un
-P_{\rm phys})}=0$ for an observable in Dirac constraining.
Moreover the proof of 5.6 used only the existence of the limit
(3.6) for $\psi$, so it will work for larger domains than $C_\varphi$.

In the particular cases where $S$ is either $C_\varphi$ or
$C_c(\r^n)$, we shorten the notation to $\cl A._\varphi:=
\cl A.\s C_\varphi.$ and $\cl A._c:= \cl A.\s C_c(\r^n).$
where $\cl A.$ can be $\cl F.,\;\cl O.,\;\cl D.$ or $\cl R.$.
Note that 
$$\eqalignno{C_c(\r^n)\cap\ker\kappa&=
\set\psi\in C_c(\r^n),\psi\rest\Gamma=0.=C_c(\r^n\backslash\Gamma)\cr
\cl D._c&=\set A\in\cl O._c,AC_c(\r^n)\subseteq C_c(\r^n\backslash\Gamma)
\supseteq A^*C_c(\r^n).\cr}$$
and for an $A\in\cl O._c$ we have $\L A'\kappa(\psi)=\kappa(A\psi)
=(A\psi)\rest\Gamma$ for all $\psi\in C_c(\r^n)$.
Analogous statements hold if we replace $C_c(\r^n)$ by $C_\varphi$.

On comparing the method above to that of Landsman [La], we note that 
the algebra which Landsman selects to impose the constraint on is
the subalgebra of the commutant $\hat\varphi'$ which preserves
$C_c(\r^n)$, i.e. $\hat\varphi'\cap\cl F._c$. In our case the
algebra which we constrain, $\cl O._c$ can be considerably
larger than that, with consequently larger algebra of observables
$\cl R._c$, but on the other hand there are also nonzero
$A\in\hat\varphi'\cap\cl F._c\backslash\cl O._c$ (see below). \chop


\noindent Next we wish to examine whether particular classes of operators
are in $\cl O._S$.
\item{$\bullet$} Consider the multiplication operator 
$\left(T_f\psi\right)(\b x'):=f(\b x')\cdot\psi(\b x')$
$\forall\,\psi\in\cl H.$ where $f$ is bounded and Borel. It is not automatic
that $T_f\in\cl F._S$. In fact if $S=C_c(\r^n)$, then
$T_fC_c(\r^n)\subseteq C_c(\r^n)$ iff $f$ is continuous
(in which case $T_f\in\cl O._c$ by 5.5).
For the choice $S=C_\varphi$ we see that if a Borel
function $f$
is discontinuous on $\Gamma$, then restriction to $\Gamma$ may not be defined,
i.e. we may have $T_f\not\in\cl F._\varphi$.
(This also shows $\hat\varphi'\backslash\cl F._\varphi\not=\emptyset$
hence that $\hat\varphi'\not\subset\cl O._\varphi\cup\cl O._c$).
On the other hand, if $f$ is continuous and bounded on some shell $S_t$, we have
$T_f\in\cl F._\varphi$. So $\cl F._\varphi$ contains a larger
class of multiplication operators than $\cl F._c$, and these
are all in $\cl O._\varphi$. In particular,
if $\exp i\hat{\b q'}\cdot\b a'\in\cl F.$ then $\exp i\hat{\b q'}\cdot
\b a'\in
\cl O._c\cap\cl O._\varphi$. However a $T_f$ is only in $\cl F._{\cl T.^{(t)}}$
if $f\in{\cl T.^{(t)}}$.
\item{$\bullet$} 
Now consider the unitaries $V_\beta$, $\beta\in{\rm Diff}\,\r^n$.
For the choice $S=C_c(\r^n)$, we have that both
$V_\beta$ and $V_\beta^*$ preserve $C_c(\r^n)$, hence
$V_\beta\in\cl F._c$ for all $\beta\in{\rm Diff}\,\r^n$. This is not
however true for the choice $S=C_\varphi$. In fact, let
$\beta$ be a fixed translation $\beta\b x'=\b x'+\b a'$ 
not preserving $\Gamma$, and let
$\psi\in C_\varphi$ be continuous on a neighbourhood of $\Gamma$
but so discontinuous on $\beta\Gamma$ that restriction to
$\beta\Gamma$ is not defined.
Then $V_{\beta^{-1}}\psi\not\in C_\varphi$, and so
$V_{\beta^{-1}}\not\in\cl F._\varphi$.\chop
Thus $\cl F._\varphi$ contains a smaller set of the unitaries
$V_\beta$ than $\cl F._c$.
Below we will examine when these are in $\cl O._\varphi$.
\item{$\bullet$} Let $S=C_c(\r^n)\ni\psi$, then 
$$\L V_\beta'\kappa(\psi)=
\kappa(V_\beta\psi)=(V_\beta\psi)\rest\Gamma=(J_\beta^{1/2}\rest\Gamma)
\cdot(\psi\circ\beta)\rest\Gamma\,.$$
So $\overline{\L V_\beta'}=\un$ iff $\beta\b x'=\b x'$ $\forall\,\b x'\in\Gamma$
and $J_\beta\rest\Gamma=1$, i.e. $V_\beta-\un\in\cl D._c$
iff $\beta\b x'=\b x'$ $\forall\,\b x'\in\Gamma$ and $J_\beta\rest\Gamma=1$.

\noindent Since $\kappa(\psi)=\psi\rest\Gamma$ for $\psi\in C_c(\r^n)$,
one might surmise that $V_\beta\Gamma\subseteq\Gamma$ iff 
$V_\beta\in\cl O._c$. However,
the example (5.3) shows that
it is not true that $\beta\Gamma\subseteq\Gamma\Rightarrow
V_\beta\in\cl O._c$. For the converse, we have:
\thrm Lemma 5.7." If $V_\beta\in\cl O._c\cup\cl O._\varphi$,
then $\beta\Gamma\subseteq\Gamma$."
Let $V_\beta\in\cl O._c$, hence $V_\beta$ preserves
$\ker(\cdot,\cdot)_\Gamma\cap C_c(\r^n)=C_c(\r^n\backslash\Gamma)$.
This is equivalent to $\beta\big(\r^n\backslash\Gamma\big)\subseteq
\r^n\backslash\Gamma$, which is equivalent to
$\beta\Gamma\subseteq\Gamma$.
This argument directly adapts to $C_\varphi$.

Note that if $\Gamma$ is curved and nonperiodic,
 then the translations will not preserve it,
so $\exp(i\hat{\b p'}\cdot\b a')\not\in \cl O._c\cup\cl O._\varphi$
hence of the generating unitaries of $\overline{\Delta(\r^{2n})}$,
only the commutative set $\set\exp(i\hat{\b q'}\cdot\b a'),\b a'\in\r^n.$
is in $\cl O._c\cup\cl O._\varphi$. This is why we consider the CCR--algebra
as too small a choice of field algebra for this type of constraining.
\def\J{\Bbb{J}}
\thrm Theorem 5.8." Given notation above, we have:\chop
$\qquad V_\beta\in\cl O._c\cup\cl O._\varphi$ 
iff $\beta\Gamma\subseteq\Gamma$ and
$J_\beta(\b x')=\J_\beta(\b x')$ for all $\b x'\in\Gamma$\chop
where $J_\beta$ (resp. $\J_\beta$) is the Jacobian of $\beta$
(resp. $\beta\rest\Gamma$) with respect to $\mu$ (resp. $\gamma$)."
Let $V_\beta\in\cl O._c\cup\cl O._\varphi$, 
so by 5.7 $\beta\Gamma\subseteq\Gamma$.
Now since $\overline{\L V_\beta^*'}=\L V_\beta'^*$ we get
$$\eqalignno{\L V_\beta'^*&\L V_\beta'\kappa(\psi)
=\L V_\beta'^*\kappa(V_\beta\psi)\cr
&=\L V_\beta^*'\kappa(V_\beta\psi)=\kappa(V_\beta^*V_\beta\psi)\cr
&=\kappa(\psi)=\L V_\beta'\L V_\beta'^*\kappa(\psi)\quad\forall\,\psi\in
C_c(\r^n)\quad\hbox{or}\quad C_\varphi\,.\cr}$$
Thus $\overline{\L V_\beta'}$ is unitary, so
$$\eqalignno{
(\psi,\,\psi)_\Gamma&=(V_\beta\psi,\, V_\beta\psi)_\Gamma
=\big(V_\beta\psi\rest\Gamma,\, V_\beta\psi\rest\Gamma\big)_{L^2(\Gamma)}\cr
&=\int_\Gamma\big|(V_\beta\psi)(\b x')\big|^2d\gamma(\b x')=
\int_\Gamma J_\beta(\b x')\cdot\big|\psi(\beta\b x')\big|^2d\gamma(\b x')\cr
&=\int_\Gamma J_\beta(\b x')\cdot\J_\beta^{-1}(\b x')
\cdot\big|\psi(\beta\b x')\big|^2
\,d\gamma(\beta\b x')\cr
&=\int_\Gamma J_\beta(\beta^{-1}\b x')\cdot\J_\beta^{-1}(\beta^{-1}\b x')
\cdot\big|\psi(\b x')\big|^2\, d\gamma(\b x')\cr
&=\int_\Gamma\big|\psi(\b x')\big|^2\, d\gamma(\b x')\qquad
\forall\,\psi\in C_c(\r^n)\quad\hbox{or}\quad C_\varphi\cr}$$
iff $J_\beta(\beta^{-1}\b x')\cdot\J_\beta^{-1}(\beta^{-1}\b x')=1$ for all
$\b x'\in\Gamma$, i.e. $J_\beta(\b x')=\J_\beta(\b x')$ for all
$\b x'\in\Gamma$.\chop
Conversely, observe that $\beta\Gamma\subseteq\Gamma$ guarantees that
both $\L V_\beta'$ and $\L V_\beta^*'$ exist, so by the reversibility
of the previous calculation when $J_\beta=\J_\beta$ on $\Gamma$,
we see that $\overline{\L V_\beta'}$ is unitary, hence bounded,
and similarly the same is true for $\overline{\L V_\beta^*'}$.
Since $\Lambda$ is a homomorphism and inverses are unique,
we get that $\overline{\L V_\beta^*'}=\L V_\beta'^*$,
i.e. $V_\beta\in\cl O._c\cup\cl O._\varphi$.

\item{\b Remarks:'}
\b(1)' Note that if $\beta\in{\rm Diff}\,\r^n$ preserves all the level sets
of $\varphi$ we have $\varphi\circ\beta=\varphi$, so
$[V_\beta,\,\hat\varphi]=0$. However since we need not have 
${\Bbb J}_\beta=J_\beta$ on $\Gamma$, there are certainly
such $\beta$ for which $V_\beta\not\in\cl O._c$.
Thus $\cl F._c\cap\hat\varphi'\backslash\cl O._c\not=\emptyset$,
i.e. Landsman [La] quantizes some operators which we have excluded from
our observables. (Some of these can still be taken
through $\Lambda$ using 5.6, but 
they need not preserve the adjoint or boundedness).
Nevertheless, we conclude that the field algebra 
$C_b(\r^n)\rtimes{\rm Diff}\,\r^n$
has ample elements in its physical algebras $\cl R._c$ or $\cl R._\varphi$.
\item{\b(2)'} There is no reason in general to expect that the time 
evolutions will preserve $\cl R._S$, so we will need to extend it by
taking the constrained field algebra as the C*-algebra generated in
$\cl B.(L^2(\Gamma))$ by $\cl R._S$ 
and the unitaries $\exp it\overline H_\Gamma$,
$t\in\r$.
\item{\b(3)'} Recall that we consider the enforcement of secondary 
quantum constraints as the selection of the set $S=\cl T.^{(t_0)}$
which the observables should preserve. Since this is an analogy to
the classical procedure, it is natural to look for reasons to justify
such a choice, and a first attempt may be to ask whether the 
choice of transverse states for $S$ will make the Jacobian condition in
5.8 obsolete, i.e. whether for a $V_\beta\in\cl F.\s{\cl T.^{(t)}}.$
we have 
$V_\beta\in\cl O.\s{\cl T.^{(t)}}.$ iff $\beta\Gamma\subset\Gamma$.
This is not true, as we can see by an easy counterexample. 
Continue the example (5.3) with the additional assumption that
the Hamiltonian has ${\rm Dom}\, H=C_c^\infty(\r^2)$ and that
$\beta\in{\rm Diff}\,\r^2$ is not the given one in (5.3), but
$\beta(x_1,\, x_2)=(x_1,\, 2x_2)$. Then both $V_\beta$ and
$V_\beta^*$ preserve
$$\cl T.^{(t_0)}=\set\psi\in C_c^\infty(\r^n),
{\partial\psi\over\partial x_2}(\b x')
=0\;\;\;\forall\;\b x'\in S_{t_\psi},\;\; t_\psi<t_0.$$
hence $V_\beta\in\cl F.\s{\cl T.^{(t_0)}}.$, and $\beta\Gamma=\Gamma$,
$\beta\rest\Gamma=\un$, $\J_\beta=1$, $J_\beta=2$, hence
$$\Lambda(V_\beta)^*=\sqrt 2\,\un\not=
\Lambda(V_\beta^*)={1\over\sqrt 2}\un\;,$$
so $V_\beta\not\in\cl O.\s{\cl T.^{(t_0)}}.$.\chop
Another possible reason one may want to use to justify the enforcement of
 secondary quantum
constraints, is to produce a common dense invariant domain 
$\kappa(S)$ for the observables $\cl R._S$. However, for the choice
$S=\cl T.^{(t)}$ there is no reason why the Hamiltonian
$H_\Gamma$ should preserve $\kappa(\cl T.^{(t)})$, so it will
be unreasonable to require the observables to do so.\chop
So at this stage we fail to see why secondary quantum constraints
should be imposed on the observables.  
\item{\b(4)'} It is interesting to observe that for any Hilbert
space operator $K:\cl H._1\to\cl H._2$ with dense range, we can obtain the
structure above. That is, given a dense space $S\subseteq{\rm dom}\, K$
for which $K(S)$ is dense in $\cl H._2$, we can define 
$\cl F._S,\;\cl O._S,\;\cl D._S,\;\cl R._S$ exactly as before by replacing
$\cl F.$ with $\cl B.(\cl H._1)$, and $\kappa$ by $K$.
So every such operator $K$ and space $S$ defines a lifting problem,
hence a short exact sequence of *--algebras
$$0\rightarrow\cl D._S\rightarrow\cl O._S\rightarrow\cl R._S\rightarrow 0\;.$$
This then produces a similar short exact sequence for the automorphism
groups and their lifting through the factorisation.
In particular, for ordinary Dirac constraining
the operator
$K$ is the projection $P_{\rm phys}:\cl H.\to\cl H._{\rm phys}$, the 
space $S$ is $\cl H.$ and $\cl A._S\cap\cl F._S$ is just $\cl A.$
where $\cl A.$ denotes either $\cl O.$ or $\cl D.$.
This way of thinking nicely unifies the current construction
(where $K=\kappa$) with that of the Dirac approach.
The main difference is that $P_{\rm phys}$ is bounded whilst
$\kappa$ is unbounded and nonclosable.

\beginsection 6. Constraining by a general selfadjoint operator.

For later reference we start by summarizing the constraining
algorithm developed in the preceding sections.
Given the following data:
operators $\hat\b p'$, $\hat\b q'$ and $H$ on $C_c^\infty(\r^n)
\subset L^2(\r^n)$, a bounded $\varphi\in C^\infty(\r^n)$
such that $\Gamma=\varphi^{-1}(0)$ is a smooth $(n-1)
\hbox{--dimensional}$ submanifold and moreover
$\nabla\varphi$ is bounded and nonzero on a neighbourhood
of $\Gamma$
and a unital field algebra $\cl F.\subset\cl B.(L^2(\r^n))$
containing $\hat\varphi$, $\exp i\hat\b p'\cdot\b a'$,
$\exp i\hat\b q'\cdot\b a'$,
\item{(1)} define $\qquad(\psi_1,\,\psi_2)_\Gamma:=
\lim\limits_{t\to 0}{1\over 2t}\big(\psi_1,\,
\hat hP_t\psi_2\big)\qquad\forall\psi_i\in C_\varphi(\r^n),$\chop
$L^2(\Gamma)=\left[C_\varphi(\r^n)\big/\ker(\cdot,\cdot)_\Gamma
\right]^{-\!-(\cdot,\cdot)_\Gamma}$,
$\kappa:C_\varphi\to L^2(\Gamma)$ is the factorisation map
and $L^2(\Gamma)$ is the constrained space.
\item{(2)} Let $\cl T._{t_0}:=\cl H._{t_0}^\zeta\cap C_\varphi\cap
{\rm Dom}\, H$, check that $H\cl T._{t_0}\subset C_\varphi$
for $t_0$ small enough, and if so, define the constrained
Hamiltonian $H_\Gamma$ by $H_\Gamma\cdot\kappa(\psi):=
\kappa(H\psi)$ for all $\psi\in\cl T._{t_0}$.
\item{(3)} Choose a space $S$ in $\cl H.$ for which $\kappa(S)$
is dense (e.g. $C_\varphi$), and define
$$\eqalignno{\cl F._S&=\set A\in\cl F.,AS\subseteq S\supseteq A^*S.\cr
\cl O._S&=\set A\in\cl F._S,(A\psi,\, A\psi)_\Gamma\leq M_A(\psi,\,\psi)_\Gamma
\geq (A^*\psi,\, A^*\psi)_\Gamma,\;\Lambda(A^*)\subseteq\Lambda(A)^*.\cr}$$
where $\Lambda(A)$ denotes the closure of the operator defined by
$$\Lambda(A)\kappa(\psi)=\kappa(A\psi)\qquad\forall\;\psi\in S\,.$$
Define $\cl D._S:=\set A\in\cl O._S,AS\subseteq\ker\kappa\supseteq
A^*S.$, then\chop
$\Lambda(\cl O._S)\cong\cl O._S\big/\cl D._S=:\cl R._S$,
and the algebra of the constrained observables on $L^2(\Gamma)$
is the C*--algebra generated by $\cl R._S$ and $\exp(i\r H_\Gamma)$.

\noindent In this section we want to 
generalise the method above to impose a constraint ``$C\psi=0$''
where $C$ is a general bounded selfadjoint operator on a Hilbert space 
$\cl H.$ with zero in its continuous spectrum. (If we start with an
unbounded selfadjoint operator $C$ we can, without loss of generality
convert the problem to a bounded one by replacing $C$ with $f(C)$
where $f$ is a continuous bounded real--valued function with
$f(x)=x$ on a neighbourhood of zero).
As before, we still assume that there is a unital field algebra
$\cl F.$ acting on $\cl H.$, containing $C$, and that there is also
a (possibly unbounded) Hamiltonian $H$ given on $\cl H.$,
$\exp(itH)\in\cl F.$ for all $t$.
We will be concerned with the construction of three objects;--
the constrained Hilbert space (and the constraining map to it from the
original space), the constrained Hamiltonian, and the algebra of constrained
observables.

At the abstract C*--level, the kinematics part has already been
solved [GH], but one obtained the set of all representations
in which the constraint can be imposed as an eigenvalue condition.
Here we  want to build a particular concrete
constrained system out of the given unconstrained one.

Now geometry was paramount in the analysis of the previous
sections, in that we needed a metric (on an underlying space)
 to define gradients, norms
of vectors, normals to surfaces and the Lebesgue 
measure. This vital piece of information
is missing in the problem under consideration, and we somehow need to
augment the given data $\{\cl F.,\,\cl H.,\,H,\, C\}$ in order to
adapt the method previously found to this problem.
The extra information we will assume is:
\item{$(\rn1)$} a maximally commutative C*--algebra $\cl A.
\subset\cl B.(\cl H.)$ containing $C$.\chop
(Call $\cl A.$ a {\it polarisation}, and we know that it always has
a cyclic and separating vector [BR 2.5.3])
\item{$(\rn2)$} A choice of a cyclic and separating vector $\Omega$
for $\cl A.$. (Call $\Omega$ the {\it vacuum}).
\item{$(\rn3)$} A choice of scaling operator $K\in\cl A._+$,
$\ker K=\{0\}$.
\item{$(\rn4)$} A selfadjoint operator $\cl P._C$ such that 
$[\cl P._C,\,C]\psi=i\psi$ for all $\psi\in\cl D.\equiv{}$ a dense invariant
subspace of $P\s t_0.\cl H.$ where $P\s t.$ denotes the
spectral projection of $C$ of the interval $[-t,\,t]$. 
This is thought of as a ``local'' canonical momentum for $C$,
which will define the normal direction to the constrained system.
\par\noindent
Now define the space $\cl L.:=\set A\in\cl A.,\omega_0(A)\;\;
\hbox{exists}\,.$ where
$$\omega_0(A):=\lim_{t\to 0}{(P_t\Omega,\, AP_t\Omega)\over\|P_t\Omega\|^2}\;.$$
Since obviously $\un$ and $P_t\in\cl L.$, this space is not zero.
Moreover, the space is selfadjoint, and we also have 
that $|\omega_0(A)|\leq\| A\|$ for all $A\in\cl L.$
because $\big|(P_t\Omega,\, AP_t\Omega)\big/\|P_t\Omega\|^2\big|\leq\|A\|$,
and so $\omega_0$ extends to the closure of $\cl L.$.
Now since $\omega_0$ is obviously positive on the positive elements
of $\cl L.$, we can apply [KR] 4.3.13 to conclude that
$\omega_0$ extends to a state on $\cl A.$. Henceforth we
fix a choice of extension and still denote it by $\omega_0$.

Denote the GNS--representation of $\omega_0$ by $(\pi_0,\,\Omega_0,\,
\cl H._0)$.
Noting that $\omega_0(K)>0$, we define a state $\omega_K$ on $\cl A.$
by
$$\omega_K(A):=\omega_0(KA)\big/\omega_0(K)\;,\qquad A\in\cl A.\,.$$
Then, inspired by the following lemma and subsequent example,
we identify the constrained Hilbert space with $\cl H.\s\omega_K.$.
\thrm Lemma 6.1."$\omega_K$ is a ``Dirac state'' in the sense that
$\omega_K(AC)=0$ for all $A\in\cl A.$, i.e. $\pi\s\omega_K.(C)
\Omega\s\omega_K.=0$."
It suffices to show that $\omega_K(C^2)=0$, since $C$ is selfadjoint.
So
$$\leqalignno{&{\omega_0(KC^2)\over\omega_0(K)}=
{\big(A^*\Omega,\, KP_tC\Omega\Big)\over
\|P_t\Omega\|^2\,\omega_0(K)}\,.\cr
\big|(KP_t\Omega,&\, P_tC^2\Omega)\big|\leq
\|KC\|\cdot\|P_t\Omega\|\cdot\|P_tC\Omega\|&\hbox{Now}\cr
&=\|KC\|\cdot\|P_t\Omega\|\cdot\|\int^t_{-t}\lambda\, dP(\lambda)\Omega\|\cr
&\leq\|KC\|\cdot\|P_t\Omega\|\cdot t\|\int^t_{-t}dP(\lambda)\Omega\|\cr
&=t\|KC\|\cdot\|P_t\Omega\|^2\;.\cr}$$
Hence$\big|\omega_K(C^2)\big|\leq{1\over\omega_0(K)}\lim\limits_{t\to 0}
t\|KC\|=0$.

\item{\bf Remark:} The proof above is easily adapted to show that
$\omega_0$ is also a Dirac state on $\cl A.$. More precisely, we
have for the left kernels 
that $N_{\omega_0}=N_{\omega_K}$ due to $\ker K=\{0\}$
and the fact that $\cl A.$ is commutative.
Hence we have an identification map $\theta:\cl A.\big/N_{\omega_0}
\to\cl A.\big/N_{\omega_K}$ by $\theta(A+N_{\omega_0})=A+N_{\omega_K}$
and $\theta$ extends to a map $\theta:\cl H._{\omega_0}\to\cl H._{\omega_K}$
because \def\slf{{1/2}}
$$\eqalignno{\big\|\theta(A+N_{\omega_0})\big\|\s{\cl H._{\omega_K}}.
&=\omega_K\big(\theta(A+N_{\omega_0})^*\,\theta(A+N_{\omega_0})\big)^\slf\cr
&=\omega_K(A^*A)^\slf=\omega_0(KA^*A)^\slf\cr
&\leq\|K\|\cdot\omega_0(A^*A)^\slf=\|K\|\cdot\|A+N_{\omega_0}\|\s{\cl H._0}..
\cr}$$
\par\noindent To define a constraining map
$\kappa:\cl H.\to\cl H.\s\omega_K.$ from the unconstrained to the constrained 
space, recall that $\Omega$ separates $\cl A.$, hence the map
$$\kappa(A\Omega):=\pi\s\omega_K.(A)\Omega\s\omega_K.\;,\qquad
A\in\cl A.$$
is well--defined on the dense subspace $\cl A.\Omega$
with dense range. This will be our choice of constraining map for
this context. Since $\|\kappa(A\Omega)\|^2=\omega_K(A^*A)$,
we see that $\ker\kappa=N_{\omega_K}\Omega=N_{\omega_0}\Omega$.
\item{\bf Example.} To motivate the preceding structures, and to see
what is involved in the choices of $\cl A.,\;\Omega,\; K,$ we now reconsider
the constraint situation $\hat\varphi$, $L^2(\r^n)$ of the preceding
sections.\chop
Starting with $\hat\varphi=\varphi(\hat\b q')$, the natural choice of 
polarisation is $\cl A.:=\set f(\hat\b q'),f\in L^\infty(\r^n).=
\set\exp(i\hat\b q'\cdot\b a'),\b a'\in\r^n.''$, and for $\Omega$
we can then choose any positive nowhere vanishing $L^2\hbox{--function}$.
Choose the Gaussian $\Omega(\b x')=\exp(-a|\b x'|^2)$, $a>0$ and note
that $C_c(\r^n)=C_c(\r^n)\Omega$, hence $C_c(\r^n)\subset\cl A.\Omega$.
Now for $K$ we must have $K\in\cl A._+$, so $K=k(\hat\b q')$
for some $k\in L_+^\infty(\r^n)$, $k^{-1}(0)=\emptyset$. For the moment
we will choose $k$ to be continuous, and then below deduce the
precise choice which will correspond with the previous results for
this situation. In particular, what we want to show is that
$$\big(\kappa(A\Omega),\,\kappa(B\Omega)\big)\s{\cl H._{\omega_K}}.
=b\big((A\Omega\rest\Gamma),\,(B\Omega\rest\Gamma)
\big)\s L^2(\Gamma).\eqno{(*)}$$
for all $A,\; B\in C_c(\r^n)=\cl L.$ 
for the right choice of $K$, where $\kappa$
is the map defined in this section and $b$ is a normalising constant. Now
$$\big(\kappa(A\Omega),\,\kappa(B\Omega)\big)\s{\cl H._{\omega_K}}.
=\omega_K(A^*B)=\lim_{t\to 0}{\big(A\Omega,\, KP_tB\Omega\big)\over
\|P_t\Omega\|^2\;\omega_0(K)}$$
whenever $A^*B\in\cl L.$ (which we will see below is all $C_c(\r^n)$)
so if we write $A=f_A(\hat\b q'),\;\; B=f_B(\hat\b q')$ with
$f_A,\; f_B\in C_c(\r^n)$ and remember that 
$$\big(P_t\psi\big)(\b x')=\chi\s\varphi^{-1}[-t,\, t].(\b x')\cdot
\psi(\b x')\;,$$
we have 
$$
{\big(A\Omega,\, KP_tB\Omega\big)\over
\|P_t\Omega\|^2}={\int_{\varphi^{-1}[-t,\, t]}\overline{f}_A(
\b x')\,k(\b x')\,f_B(\b x')\,\exp(-2a|\b x'|^2)\;d\mu(\b x')\over
\int_{\varphi^{-1}[-t,\, t]}\exp(-2a|\b x'|^2)\;d\mu(\b x')}$$
Now recall that for $f\in C_c(\r^n)$ and small $t$ we have
$$\leqalignno{\int_{\varphi^{-1}[-t,\, t]}f(\b x')\; d\mu(\b x')
&=\int_{-t}^t\Big(\int{f(\b y',\,\varphi)
\over\big|\nabla\varphi(\b y',\,\varphi)
\big|}\,d\gamma_\varphi(\b y')\Big)\; d\varphi\cr
&=2t\int{f(\b y',\, 0)\over\big|\nabla(\b y',\, 0)\big|}\; d\gamma(\b y') 
+O(t)\cr
&=2t\int_\Gamma\big(f\Big/|\nabla\varphi|\big)\rest\Gamma\; d\gamma + O(t)\;.\cr
\big(\kappa(A\Omega),\,\kappa(B\Omega)\big)\s{\cl H._{\omega_K}}.
&=\lim_{t\to 0}{\big(A\Omega,\, KP_tB\Omega\big)\over
\|P_t\Omega\|^2\omega_0(K)}&\hbox{thus:}\cr
&={\int_\Gamma\Big(\overline{f}_Akf_Be\big/|\nabla\varphi|
\Big)\rest\Gamma\;d\gamma
\over\omega_0(K)\int_\Gamma\big(e\big/|\nabla\varphi|\big)\rest\Gamma\; d\gamma}
\cr}$$
where $e$ denotes the function $\exp(-2a|\b x'|^2)$. Now since
by a similar argument
$$\omega_0(K)=\lim_{t\to 0}{\big(P_t\Omega,\, KP_t\Omega\big)\over\|
P_t\Omega\|^2}={\int_\Gamma\big(ke\big/|\nabla\varphi|\big)\rest\Gamma\;
d\gamma\over\int_\Gamma\big(e\big/|\nabla\varphi|\big)\rest\Gamma\; d\gamma}$$
(the existence of this limit shows $\cl L.\supseteq C_c(\r^n)$) we have $$\leqalignno{
\big(\kappa(A\Omega),\,\kappa(B\Omega)\big)\s{\cl H._{\omega_K}}.
&={\int_\Gamma\Big(\overline{f}_A
kf_Be\big/|\nabla\varphi|\Big)\rest\Gamma\;d\gamma
\over\int_\Gamma\big(ke\big/|\nabla\varphi|\big)\rest\Gamma\; d\gamma}
\cr
={\int_\Gamma(\overline{f}_Af_Be)\rest\Gamma\;d\gamma
\over\int_\Gamma e\rest\Gamma\; d\gamma}
&={\vphantom{\Big|}
\big((A\Omega\rest\Gamma),\,(B\Omega\rest\Gamma)\big)\s L^2(\Gamma).
\over\|\Omega\rest\Gamma\|^2\s L^2(\Gamma).}\cr}$$
(where we made the choice $k=|\nabla\varphi|$), 
which is just the normalised inner product
of $L^2(\Gamma)$, and this will produce the same $\kappa$ than we had before.

\noindent
 If there is also a group of symmetries which
needs to be unitarily implemented on the constrained Hilbert space
$\cl H._{\omega_K}$, we can use this unitarity to determine $K$.

So now that we have the constraining map $\kappa:\cl H.\to\cl H.\s\omega_K.$
as above, we can proceed to constrain the dynamics. This is where we will
use the assumed operator $\cl P._C$ in the role of the ``normal derivative $i
{\partial\over\partial\varphi}$ near $\Gamma$.'' Thus start by selecting
a space of transverse states by
$$\leqalignno{\cl H._t^T&:=\overline{\set  P_t\psi,{\psi\in\dom \cl P._C\,,\;\;
P_t\cl P._C\psi=0}.}\cr
\cl T._t&:=\set\psi\in\dom H,P_t\psi\in{\cl H._t^T}.&\hbox{and}\cr}$$
and note that $\cl T._t\subset\cl T._s$ if $t>s$.
For a consistent constraining we need to assume there is a $t_0>0$
such that the set $\cl S._{t_0}:=\set\psi\in\cl T._{t_0},
H\psi\in{\cl A.}\Omega.$ satisfies $\ker\kappa\cap\cl S._{t_0}=\{0\}$
and $\kappa(\cl S._{t_0})$ is dense in $\cl H._{\omega_K}$.
Given this, we define the constrained Hamiltonian $H_C$ on domain
$\kappa(\cl S._{t_0})$ to be
$$H_C\cdot\kappa(\psi):=\kappa(H\psi)\;\;\forall\;\psi\in\cl S._{t_0}\,.$$
\itemitem{\bf Example.} Continue the previous example with $C=\hat\varphi$
and choice $K=|\nabla\varphi|$. Choose
$$\cl P.\s\hat\varphi.\psi:=i\,
 f(\varphi)\,{\partial\over\partial\varphi}\,\psi
\qquad\forall\psi\in C_c^\infty(\r^n)$$
where $f$ is a smooth bump function which is one on $[-t_0,\, t_0]$
and zero outside $[-t_0-\varepsilon,\, t_0+\varepsilon]$ for some
$\varepsilon>0$. Then \def\supp{{\rm supp}\,}
$$\eqalignno{\cl H.\s t_0.^T=\big\{\psi\in\cl H.\;&\big|
\;\supp\psi\subset\varphi^{-1}\big([-t_0,\,
t_0]\big)\,,\;\;\psi(\b y',\varphi) \;\;\hbox{independent of $\varphi$}\cr
&\qquad\qquad\qquad\qquad\qquad\hbox{ on
$\varphi^{-1}[-t_0,t_0]$}\;\big\}\cr}$$
If $H$ is a differential operator with domain $C_c^\infty(\r^n)$, we have
$$\eqalignno{
\cl T._{t_0}=\set\psi\in C_c^\infty(\r^n), \psi\rest\varphi^{-1}((-t_0,\,
t_0))\;\;\hbox{is constant in the direction}\;\;\nabla\varphi..\cr}$$
Since $\cl H.\s t_0.^\zeta\cap C_c^\infty(\r^n)\subset\cl T._{t_0}$,
we conclude that
$$H\s\hat\varphi.\kappa(\psi)=\kappa(H\psi)\quad\forall\psi\in\cl T.\s t_0.$$
defines the same operator as before.

Finally, to obtain the constrained kinematics as in Sect. 4, let
$S\subseteq\cl A.\Omega$ be a subspace such that $\kappa(S)$ is
dense in $\cl H._{\omega_K}$. Define in complete analogy with Sect. 
4, $$\leqalignno{\cl F._S&:=\set A\in\cl F.,A\psi\in S\ni A^*\psi
\quad\forall\;\psi\in S.\cr
\cl O._S&:=\big\{\; A\in\cl F._S\;\big|\;(A\psi,\, A\psi)_K
\leq M_A\cdot(\psi,\,\psi)_K\geq (A^*\psi,\,A^*\psi)_K\;\;
\forall\psi\in S,\cr 
&\qquad\qquad\qquad\qquad\qquad M_A<\infty\quad
\hbox{and}\quad\Lambda(A^*)\subseteq\Lambda(A)^*\;\big\}\cr
\cl D._S&:=\set A\in\cl O._S,AS\subseteq\ker\kappa\supseteq A^*S.\cr
&\Lambda(A)\,\kappa(\psi):=\kappa(A\psi)\;\forall\,\psi\in S,\;
A\in\cl O._S\;.&\hbox{where}\cr}$$
The previous proofs carry over verbatim, so $\cl D._S=\ker\big(\Lambda
\rest\cl O._S\big)$ and the algebra of observables acting on
$\cl H._{\omega_K}$ is the C*--algebra generated by
$\exp i\r H_C$ and
$$\cl R._S:=\cl O._S\big/\cl D._S\cong \Lambda(\cl O._S)
\subset\cl B.(\cl H._{\omega_K})\;.$$
\itemitem{\bf Example.} In continuation of the last example, it is obvious
that if we take $\cl F.=C_b(\r^n)\rtimes{\rm Diff}\, \r^n$ and
$S=C_c(\r^n)$, then we will get the same observable algebra
than before.
\itemitem{\bf Remark} Another way to proceed could have been to
extend the Dirac state $\omega_K$ on $\cl A.$ to $\cl F.$.
However, this would have enlarged the GNS--space (producing 
a wrong result for the example of this paper), and in addition
there is the problem that the extension need not be unique.
\beginsection 7. Conclusions and Discussion.

To summarize;- we took the problem of restricting a quantum particle
in $\r^n$ to a lower dimensional submanifold not conserved by the dynamics.
We found a method to do the restriction, and this came in three parts;-
a map $\kappa$ from a dense subspace of $L^2(\r^n)$ to the constrained
Hilbert space $L^2(\Gamma)$, a method 
for constructing the constrained Hamiltonian
from the unconstrained one, and a method for the construction of the algebra
of observables $\cl R._\Gamma$ on $L^2(\Gamma)$.
For each of these three parts there were choices to be made;- for the first part
a domain for $\kappa$ ($C_c(\r^n)$ and $C_\varphi$ seem the canonical 
choices), for the second part a choice of transverse space (here
tangentiality of momentum 
to the level sets of $\varphi$ near $\Gamma$ seems to be the physically
justified criterion), and for the 
third part a choice of space $S$ on which to reduce
the observables (likely choices here are $C_c(\r^n)$, $C_\varphi$
or the transverse space used). We regard these choices to be made as
physical ones, and in the two examples, the choices made were evident, and
produced results in agreement with the known solutions.
The necessity of these choices should be compared with the
choice  of boundary conditions required in the constraining 
of a free quantum particle to a box.

There are mathematical pathologies associated with our proposed method, of which
the central one is that the map $\kappa$ is neither bounded nor even closable.
As a consequence, the constraining of observables need not preserve boundedness
or involution, so we had to restrict the method to those observables for
which these pathologies do not occur. The method is also sensitive with
respect to the original choice of field algebra, and may not provide enough 
observables for some choices.

Whilst this method was developed in analogy with the classical constraint method,
the analogy became quite vague at certain points, for instance, one can interpret
the neccessity for a choice of transverse space as an imposition of a secondary
constraint, but this constraint could not be interpreted as obtained from
a similar method as the classical one. Moreover secondary constraints do not
seem to be neccessary for the constraining of the observables.

We expect from our work in [GH] that the pathologies which occurred here can
be circumvented by a suitable generalisation to a C*--algebra framework,
insofar the T--procedure solves already without pathology 
the kinematics part of the question
there, but this would involve losing the original representation
which may contain some physical information. The advantage of the current
method is that it constructs the constrained
system concretely from the given representation of the original system.
Nevertheless, we intend to further pursue this line of thought.
\chop
{\bf 8. Acknowledgements.}\chop
One of us (H.G.) would like to thank Alan Weinstein for his question
during a seminar, which lead to this paper. H.G. would also like to
thank and acknowledge the support of the Erwin Schr\"odinger Institute
of Mathematical Physics during the last phase of this project.
We are grateful to Klaas Landsman for drawing our attention to the work
of [Ma].

\beginsection Bibliography.

\item{[AM]} Abraham, R., Marsden, J.E.: Foundations of mechanics.
Benjamin Publishing Co., London, 1982.
\item{[Ar]} Arnold, V.I.: Mathematical methods of classical mechanics.
Springer--Verlag, New York, 1989.
\item{[BR]} Bratteli, O., Robinson, D.W.: Operator Algebras and Quantum 
Statistical Mechanics I, Springer-Verlag, New York, 1987.
\item{[Di]} Dirac, P.A.M.: Lectures on quantum mechanics.
New York: Belfer graduate school of science, Yeshiva University, 1964.
\item{[Fa]} Farquhar, I.E.: Ergodic theory in statistical mechanics.
Interscience publishers, New York, 1964.
\item{[Fu]} Fujii, K., Ogawa, N., Uchiyama, S., Chepilko,
N.: Geometrically induced gauge
structure on manifolds embedded in a higher dimensional space,
Preprint  hep--th/9702191.
\item{[GNH]} Gotay, M.J., Nester, J.M., Hinds, G.: Presymplectic 
manifolds and the Dirac--Bergman theory of constraints,
J. Math. Phys. {\bf 19}, 2388--2399 (1978)
\item{[GGT]} Gotay, M.J., Grundling, H.B., Tuynman, G.M.:
Obstruction results in quantization theory,
J. Nonlin. Sc. {\bf 6}, 469--498 (1996)
\item{[GH]} Grundling, H., Hurst, C.A.: Algebraic quantization
of systems with a gauge degeneracy, Commun. Math. Phys. {\bf 98},
369--390 (1985)\chop
Grundling, H., Hurst, C.A.: A Note on regular states and
supplementary conditions, Lett. Math. Phys. {\bf 15}, 205--212
(1988) and errata, Lett. Math. Phys. {\bf 17}, 173--174 (1989)
\item{[Gr]} Grundling, H.: BRST quantum theory: a functional
analytic approach. UNSW preprint, 1991
\item{[He]} Helgason, S.: Groups and geometric analysis, integral geometry,
invariant differential operators, and spherical functions.
Academic press, 1984.
\item{[HT]} Henneaux, M., Teitelboim, C.: Quantization of gauge systems,
Princeton University press, Princeton, 1992.
\item{[Kh]} Khinchin, A.I.: Mathematical foundations of statistical mechanics.
Dover publications, New York, 1949.
\item{[KR]} Kadison, R.V., Ringrose, J.R.: Fundamentals of the Theory
of Operator Algebras I. Academic Press, New York 1983. 
\item{[LL]} Landsman, N.P., Linden, N.: Superselection rules from Dirac and
BRST quantization of systems with constraints, Nucl. Phys. {\bf 371B}, 415--433
(1992)
\item{[La]} Landsman, N.P.: Rieffel induction as generalised quantum
Marsden--Weinstein reduction, J. Geom. Phys. {\bf 15}, 285--319 (1995)
\item{[Ma]} Maraner, P.: Monopole gauge fields and quantum potentials induced
by the geometry in simple dynamical 
systems. Ann. Phys. {\bf 246}, 325--346 (1996).
\item{[Mc]} McMullan, D.: Classical states and the BRST charge.
Commun. Math. Phys. {\bf 149}, 161--174 (1992)
\item{[MW]} Marsden, J.E., Weinstein, A.: Reduction of symplectic manifolds with
symmetry, Rep. Math. Phys. {\bf 5}, 121--130 (1974)
\item{[RS]} Reed, M., Simon, B.: Methods of modern mathematical physics I.
Academic press, New York, 1972.
\item{[Ri]} Rieffel, M.A.: Deformation 
quantization of Heisenberg manifolds. Commun.
Math. Phys. {\bf 122}, 531--562 (1989)
\item{[SW]} Strocchi, F., Wightman, A.S.: J. Math. Phys. {\bf 15}, 2198 (1974)
and J. Math. Phys. {\bf 17}, 1930 (1976)
\item{[SM]} Sudarshan, E.C.G., Mukunda, N.: Classical dynamics:
a modern perspective. John Wiley and sons, New York, 1974.
\item{[Su]} Sundermeyer, K.: Constrained dynamics. Lect. Notes Physics {\bf 169},
Springer--Verlag, New York, 1982.
\item{[To]} Tondeur, Ph.: Foliations on Riemannian manifolds.
Springer--Verlag, New York, 1988.
\item{[VH]} Van Hove, L.: Sur certaines repr\'esentations unitaires d'un groupe
infini de transformations. Proc. Roy. Acad. Sci. Belgium {\bf 26},
1--102 (1951)
\item{[Wo]} Woodhouse, N.M.J.: Geometric quantization. Clarendon press, Oxford,
1992.

\bye